\documentclass[%
 pra,superscriptaddress,
 amsmath,amssymb,
 aps,
twocolumn, 
longbibliography]{revtex4-2}
\usepackage[utf8]{inputenc}
\usepackage{graphicx}
\usepackage{dcolumn}
\usepackage{bm}
\usepackage{physics}
\usepackage{xcolor}
\usepackage{lipsum}
\usepackage[colorlinks = true, urlcolor = cyan, linkcolor=blue, citecolor=blue, filecolor=magenta]{hyperref}
\usepackage[export]{adjustbox}

\makeatletter
\renewcommand*{\@fnsymbol}[1]{\ensuremath{\ifcase#1\or \dagger\or *\or \ddagger\or
   \mathsection\or \mathparagraph\or \|\or **\or \dagger\dagger
   \or \ddagger\ddagger \else\@ctrerr\fi}}
\makeatother

\usepackage{verbatim} 

\begin{document}

\preprint{APS/123-QED}

\title{Dissipation Enables Strongly Detuning-Dependent Interference in Pulsed Dynamical Decoupling}

\author{Kenneth DeRose\@{*}}
\affiliation{Department of Physics and Astronomy, Center for Fundamental Physics, and Center for Interdisciplinary Exploration and Research in Astrophysics, Northwestern University, Evanston, Illinois, USA}
\affiliation{Fermi National Accelerator Laboratory, Batavia, Illinois, USA}

\author{Jonah Glick\@{*}}
\affiliation{Department of Physics and Astronomy, Center for Fundamental Physics, and Center for Interdisciplinary Exploration and Research in Astrophysics, Northwestern University, Evanston, Illinois, USA}
\affiliation{Fermi National Accelerator Laboratory, Batavia, Illinois, USA}

\author{Kefeng Jiang}
\affiliation{Department of Physics and Astronomy, Center for Fundamental Physics, and Center for Interdisciplinary Exploration and Research in Astrophysics, Northwestern University, Evanston, Illinois, USA}

\author{Hans Johnson}
\affiliation{Superconducting Quantum Materials and Systems (SQMS) Center, Fermi National Accelerator Laboratory, Batavia, IL 60510, USA}
\affiliation{Embedded Computing and Signal Processing (ECASP) Research Laboratory, Department of Electrical and Computer Engineering, Illinois Institute of Technology, Chicago, IL 60616, USA}

\author{Tanay Roy}
\affiliation{Superconducting Quantum Materials and Systems (SQMS) Center, Fermi National Accelerator Laboratory, Batavia, IL 60510, USA}

\author{Sharika Saraf}
\affiliation{Department of Physics and Astronomy, Center for Fundamental Physics, and Center for Interdisciplinary Exploration and Research in Astrophysics, Northwestern University, Evanston, Illinois, USA}
\affiliation{Fermi National Accelerator Laboratory, Batavia, Illinois, USA}

\author{Anya Abraham}
\affiliation{Department of Physics and Astronomy, Center for Fundamental Physics, and Center for Interdisciplinary Exploration and Research in Astrophysics, Northwestern University, Evanston, Illinois, USA}

\author{Hardeep Singh}
\affiliation{Department of Physics and Astronomy, Center for Fundamental Physics, and Center for Interdisciplinary Exploration and Research in Astrophysics, Northwestern University, Evanston, Illinois, USA}

\author{Tim Kovachy}
\affiliation{Department of Physics and Astronomy, Center for Fundamental Physics, and Center for Interdisciplinary Exploration and Research in Astrophysics, Northwestern University, Evanston, Illinois, USA}

\begin{abstract} 

One of the defining features of pulsed dynamical decoupling is its suppression of a driven qubit's sensitivity to static detuning errors between the drive field and qubit resonance. In this paper, we show that dissipation, in the form of excited-state decay, can change this behavior entirely, producing an interference signal with a strong detuning dependence. This signal arises from decay during driven evolution and relies on coherence retained by the qubit after such a decay event. We develop analytical and numerical models that capture the underlying mechanism and observe this same dissipation-induced detuning dependence experimentally in both a free-space strontium atom interferometer and a superconducting transmon qubit system. We also use this dissipation-induced detuning dependence as the basis for a new spectroscopic technique called Dissipative Carr-Purcell Spectroscopy (DCPS) and compare it with a traditional Ramsey sequence. Our results establish a regime of pulsed dynamical decoupling in which dissipation reshapes, rather than merely degrades, coherent control, and we expect these dynamics to be relevant to a wide range of quantum systems.

\end{abstract}

\maketitle

\section{Introduction}

Coherent control of quantum systems is central to quantum sensing, quantum information processing, and precision measurement. In practice, however, quantum systems are never perfectly closed, so their dynamics are shaped by the interplay between coherent driving and dissipation.
This interplay underlies a wide range of phenomena, including steady-state coherence \cite{scully1997quantum, metcalf1999laser}, dissipative state engineering \cite{diehl2008quantum, verstraete2009quantum, plenio2002entangled, wang2013reservoir, poyatos1996quantum, kienzler2015quantum}, and driven-dissipative many-body dynamics \cite{fazio2025many}. Understanding when dissipation merely degrades coherent control and when it qualitatively reshapes it remains an important open question in quantum science.

Dynamical decoupling (DD) is a powerful tool for extending coherence in quantum systems \cite{viola1999dynamical, Suter2016Colloquium, biercuk2009optimized}, with demonstrated utility across many quantum platforms, including trapped ions \cite{biercuk2009optimized}, superconducting qubits \cite{bylander2011noise}, nitrogen-vacancy centers \cite{de2010universal}, and atomic systems \cite{Sagi2010Dynamical}. These platforms can be modeled as effective two-level qubits. A central strength of DD is that it suppresses errors associated with a static offset between the frequency of the drive field and the qubit resonance (i.e., detuning errors) \cite{carr_effects_1954}. Some DD sequences work by applying multiple $\pi$ pulses that invert the sign of the detuning-dependent phase accumulated by a qubit, so that the phase error before each pulse is canceled by an equal and opposite contribution after it, causing the net detuning-induced phase to vanish by the end of the sequence. In this paper, we restrict our attention to DD sequences implemented through discrete pulses, but DD can also be performed with continuous drive fields \cite{Cai2012Continuous, Timoney2011Continuous}.

In this paper, we show that this insensitivity to detuning errors can break down in the presence of dissipation. Specifically, excited-state decay (i.e., relaxation) can cause a DD sequence that would otherwise suppress detuning errors to instead leave the qubit's final state with a strong detuning dependence.
The detuning insensitivity normally attributed to DD (in the absence of dissipation) depends on precise pulse timing. For example, the phase error accumulated before each $\pi$ pulse only cancels the phase error accumulated afterwards if the two intervals are of equal duration.
A relaxation event partway through the DD sequence can reset a qubit to $|0\rangle$, but the remaining pulses in the sequence then continue to drive this reset qubit. These remaining pulses can the induce a detuning-dependent coherence. In this effect, dissipation can reshape, rather than solely degrade, coherent control.
Notably, this dissipation-induced effect is distinct from the detuning-dependent steady-state coherence familiar from continuously driven open systems \cite{scully1997quantum, metcalf1999laser}, since it is not a steady-state effect.

\begin{figure*}
\centering
\includegraphics[width=6.8in]{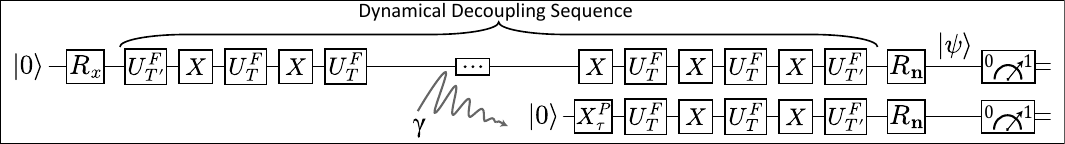}
\caption{
Gate-based picture of the impact of relaxation on the DD sequence described in Sec. \ref{sec:mechanism_for_relaxation_induced_detuning_dependence_in_DD_sequences}.
In this picture, we imagine a relaxation event (rate $\gamma$) occurs during a single $X$-gate operation, so that the qubit resets to the $|0\rangle$ state and is subject to a `partial' $X$-gate operation, $X^P_{\tau}$, and subsequently only the remaining pulses of the DD sequence. This partial $X$-gate pulse can act as a $\pi/2$ pulse rotation, splitting the $\ket{0}$ state population into a superposition of $\ket{0}$ and $\ket{1}$, which begins a new DD sequence with imperfect pulse timings. Therefore, this relaxed qubit can still contribute to the post-readout interference fringe, leading to dephasing from the coupling of detuning errors and relaxation.
}
\label{fig:all0_circuit_diagram}
\end{figure*}

We develop analytical and numerical models of this effect using a driven two-level-system description with relaxation, showing that the behavior does not require platform-specific ingredients. We observe the resulting relaxation-induced detuning sensitivity experimentally in two distinct physical platforms: a strontium (Sr) atom interferometer and a superconducting circuit hosting a transmon qubit. Despite their very different physical implementations,
both systems are well described as driven two-level systems with relaxation, providing a common framework for understanding why the same effect appears in both. These results suggest that analogous effects may arise broadly in driven quantum systems with similar relaxation dynamics.

We find that relaxation affects different DD sequences differently. The extent to which a qubit becomes sensitive to detuning via relaxation depends strongly on the particular DD pulse sequence used \cite{carr_effects_1954, gullion_new_1990, ryan_robust_2010, souza_robust_2011, genov_arbitrarily_2017}. We perform numerical simulations to characterize this relaxation-induced detuning sensitivity for several common DD sequences. We also collect experimental data on a particular sequence \cite{wang2024robust} for which relaxation destroys the initial qubit state, but produces strong coherence, even in the absence of detuning.

 We further show that this detuning dependence can be repurposed as a spectroscopic diagnostic, which we term dissipative Carr-Purcell spectroscopy (DCPS). Its underlying mechanism is qualitatively distinct from that of conventional approaches such as Ramsey spectroscopy \cite{Ramsey1950Ramsey}, giving it a different sensitivity to high-frequency detuning fluctuations, and it may find use as a complementary tool under a constrained set of experimental conditions.

 Our results establish a new perspective on pulsed DD that we anticipate will become increasingly relevant as control fidelities improve and DD sequences are extended to longer durations across different physical platforms \cite{wang2024robust, de2010universal, bar2013solid, schmitt2017submillihertz, bylander2011noise, bishof2013optical, kolkowitz2016gravitational, Graham2016_GW}.  Namely, the same extension of DD sequence time that is ordinarily used to prolong coherence \cite{viola1999dynamical, Suter2016Colloquium, biercuk2009optimized,de2010universal} or to enhance quantum-lock-in signal amplification \cite{bylander2011noise,wang2024robust, bar2013solid, de2011single, kotler2011single, boss2017quantum, maze2008nanoscale, shaniv2017quantum, schmitt2017submillihertz, shibata2021quantum, zhuang2021many, kolkowitz2012coherent,bishof2013optical, kolkowitz2016gravitational, Graham2016_GW} can also increase sensitivity to static detuning through the interplay of coherent driving and dissipation.

 The remainder of this paper is organized as follows: In Sec. \ref{sec:mechanism_for_relaxation_induced_detuning_dependence_in_DD_sequences}, we develop the physical picture and theoretical framework for relaxation-induced detuning sensitivity in pulsed DD sequences and present experimental measurements in both the atom interferometer and a superconducting qubit system. We also analyze the dissipation-enabled coherence for a range of common DD sequences. In Sec. \ref{sec:DCPS}, we introduce DCPS, analyze its response to frequency noise, and compare it with Ramsey spectroscopy. We conclude in Sec. \ref{sec:outlook} with a discussion of future directions for exploration.

\vspace{-2ex}

\section{Mechanism for Relaxation-Induced Detuning Dependence in DD Sequences}\label{sec:mechanism_for_relaxation_induced_detuning_dependence_in_DD_sequences}

Here we provide a theoretical description of the emergence of a strong detuning-dependent coherence from relaxation in certain DD sequences.
We consider an ensemble of two-level systems (e.g. an ensemble of qubits) driven by a pulsed control field in the presence of a decay channel from the excited to the ground state, with characteristic rate $\gamma = 1/T_1$, where $T_1$ is the energy relaxation time.
The ensemble of qubits can be realized by running experiments across multiple qubits simultaneously or by repeated single-qubit measurements.

We consider a DD sequence consisting solely of $X$-gate operations (sometimes referred to as a Carr-Purcell (CP) sequence \cite{carr_effects_1954}). A gate-based illustration is provided in Fig.~\ref{fig:all0_circuit_diagram}. The sequence begins with a $\pi/2$ pulse rotation about the $x$-axis ($R_{\boldsymbol{x}}$), preparing an equal superposition of ground and excited states. This is followed by a free evolution interval of duration $T^{\prime}$ (denoting the deadtime between the initial $\pi/2$ pulse and the first $\pi$ pulse), $N$ $X$-gate operations each separated by a free evolution time $T$, and a final free evolution interval of duration $T^{\prime}$. The successive $X$ gates play the role of spin echoes, flipping the qubit and decoupling it from detuning errors. The sequence concludes with a readout rotation  $R_{\boldsymbol{n}}$ (i.e., a $\pi/2$ pulse) about an axis $\boldsymbol{n} = \cos[\phi_R]\,\mathbf{x} - \sin[\phi_R]\,\mathbf{y}$, which maps the accumulated phase onto a population difference; the populations in $\ket{0}$ and $\ket{1}$ are then measured. Repeating this measurement over a range of $\phi_R \in [0, 2\pi]$ produces an interference fringe whose phase $\Delta\phi$ encodes the relative phase accumulated between the $\ket{0}$ and $\ket{1}$ states over the course of the sequence. The fringe visibility $v$ provides a measure of the magnitude of the qubit coherence immediately before the readout pulse. The duration $T^{\prime}$ is chosen such that, in the absence of relaxation and assuming error-free gate operations, $\Delta\phi$ is independent of detuning.

\begin{figure}
    \centering
    \includegraphics[scale = 1]{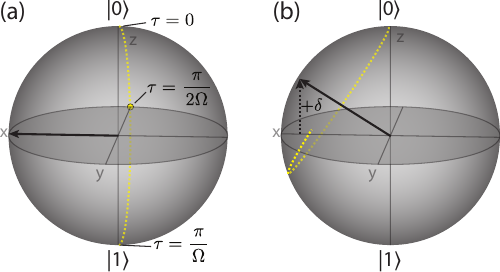}
    \caption{Two examples of an ``arc" distribution of states formed from decayed qubits during a single $X$-gate pulse in the Bloch sphere representation of a qubit in the driving field reference frame. (a) is without detuning, and (b) is with a positive detuning. The solid black arrow represents the normalized rotation vector corresponding to the gate, whose direction depends on detuning $\delta$. For the purposes of this figure, $\delta$ is taken to be normalized relative to the effective Rabi frequency $\sqrt{\Omega^2 + \delta^2}$. The yellow dotted curve represents the distribution of qubit states after a decay event to state $|0\rangle$ and a remaining pulse duration $\tau$.}
    \label{fig:CP_arc}
\end{figure}

Relaxation can cause otherwise perfect DD sequences to no longer completely preserve the intended post-readout interference between the $\ket{0}$ and $\ket{1}$ states, causing both the measured phase shift ($\Delta\phi$) and the visibility to acquire a dependence on the detuning between the drive field and the qubit.
The central mechanism is as follows: A spontaneous decay event can occur at any point during a DD sequence, at which time the qubit collapses from its prior superposition to the ground state, erasing all previously accumulated phase information. The remaining portion of the DD sequence then acts on the reset ground state.
Because qubits that decay during the DD sequences experience a truncated DD sequence that no longer cancels detuning errors, the distribution of decay times translates directly into a detuning-dependent spread in the accumulated phase, biasing the ensemble-averaged measurement of $\Delta\phi$.
This effect can be visualized on the Bloch sphere. Consider a qubit that decays during a $\pi$ pulse. As shown in Fig. \ref{fig:CP_arc}(a), the sub-ensemble of qubits that decayed at different times during the pulse therefore occupies a continuum of states tracing a smooth arc on the Bloch sphere, spanning from the south pole to the north pole. The presence of detuning modifies the radius and orientation of this arc, as shown in Fig.~\ref{fig:CP_arc}(b).

\subsection{Theory of Relaxation-Induced Detuning Sensitivity in the Carr-Purcell (CP) Sequence and Corresponding Experimental Data}\label{sec:all0_theory_intro}

We can capture the impact of the coupling of relaxation and detuning errors analytically in the following way: Consider a qubit that has undergone a relaxation event at some time during the DD sequence, as indicated in Fig. \ref{fig:all0_circuit_diagram}.
Depending on whether this event occurs during a pulse (intra-pulse decay) or during the deadtime between pulses (inter-pulse decay), the state of this qubit $|\psi\rangle$ just before it is measured can be written as a piecewise function,

\begin{equation}\label{eqn:sequenceGates}
\begin{split}
& |\psi\rangle = 
\\
&
\begin{cases}
R_{\mathbf{n}}U_{T'}^{F}
\overbrace{X U_{T}^{F} X \cdots U_{T}^{F} X}^{\text{N $X$-gates}} U_{T}^{F}X^{P}_{\tau}|0\rangle,
&
\!\!\!\! \text{intra-pulse decay}
\\
R_{\mathbf{n}}U_{T'}^{F}
\underbrace{X U_{T}^{F} X \cdots U_{T}^{F} X}_{\text{N $X$-gates}} U_{\tau}^{F}|0\rangle,
&
\!\!\!\! \text{inter-pulse decay}
\end{cases}
\end{split}
\end{equation}

\noindent
where $\tau$ denotes the partial time that the qubit is subject to immediately following the relaxation event. $U_t^{F}$ denotes free propagation, with no drive field applied, for a time $t$, and $X_{t}^{P}$ denotes a partial $X$-gate operation of duration $t$, with $t = \pi/\Omega$ corresponding to a full $X$-gate operation ($X^{P}_{\pi/\Omega} = X$).   Here, $\Omega$ is the detuning-free Rabi frequency.
At the end of the DD sequence, we read out the phase shift between the $\ket{0}$ and $\ket{1}$ states by measuring the population in the $\ket{1}$ state as a function of the phase $\phi_R$ of the final $R_{\mathbf{n}}$ readout $\pi/2$ pulse.
We can define a projection operator onto this $\ket{1}$ state as $P_1 = | 1 \rangle \langle 1 |$.
Assuming the DD sequence is long enough that every qubit in the system relaxes at least once, and averaging over all possible relaxation times as detailed in Appendix~\ref{sec:gate-based_interferometer_response_analytics}, the resulting $|1\rangle$-state population, to first order in the detuning error $\delta$, is

\begin{equation}
\begin{split}
\label{Eq:DetuningSlope}
\left\langle p_1\right\rangle &= 
\mathbb{E}_{\text{decay times}}\!\left[\langle \psi|P_1|\psi\rangle\right]
\\
&= \frac{1}{2} - \frac{2 T + \tau_{\pi}}{2\left(T+\tau_{\pi}\right)}\frac{\delta}{\Omega}\sin\left[\phi_R\right]
+\mathcal{O}\left[\delta^2\right],
\end{split}
\end{equation}

\noindent where $\mathbb{E}_{\text{decay times}}$ denotes a weighted averaging over all possible times the qubit could relax.
Expressing this population in the standard form of an
interferometric fringe, 

\begin{equation}
    \langle p_1\rangle = \frac{1}{2}\left(1 + v \cos\left[\Delta \phi
+ \phi_R\right]\right),
\label{vis_def}
\end{equation}

\noindent gives the phase shift $\Delta\phi$ and visibility $v$ to first order in $\delta$, where

\begin{equation}\label{eq:all0_v_and_phase_first_order_in_delta}
\begin{split}
v & = \frac{2 T + \tau_{\pi}}{T+\tau_{\pi}}\left|\frac{\delta}{\Omega}\right| +\mathcal{O}\left[\delta^2\right],
\\
\Delta\phi & =
\pi + 
\begin{cases}
-\frac{\pi}{2}, & \delta > 0,
\\
\frac{\pi}{2}, & \delta < 0
\end{cases}
+\mathcal{O}\left[\delta\right].
\end{split}
\end{equation}

\begin{figure}[t]
    \centering
    \includegraphics[width=3.4in]{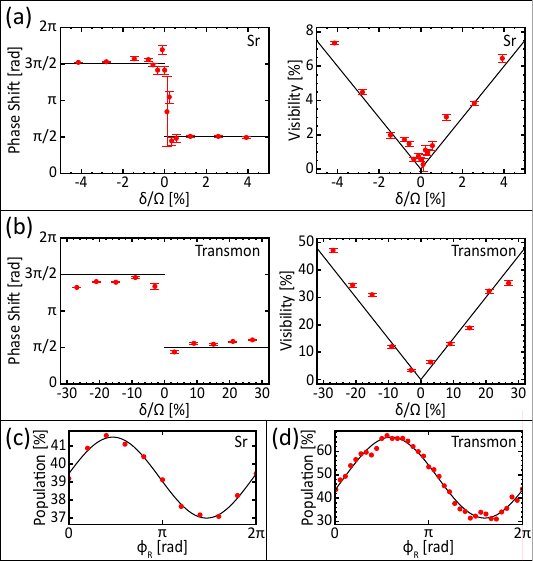}
    \caption{
    The impact of relaxation on a DD sequence in (a) a $^{88}$Sr atom interferometer with a $|1\rangle$-state lifetime $\approx21.3~\mu\text{s}$ and (b) a flux-tunable superconducting transmon qubit with $|1\rangle$-state lifetime $\approx21~\mu\text{s}$. The measured phase shift $\Delta\phi$ and visibility $v$ are observed in the regime where the total DD sequence duration exceeds the $|1\rangle$-state lifetime (504 $X$-gates, total duration $\approx113~\mu\text{s}$ in (a), and 200 $X$-gates, total duration $\approx120~\mu\text{s}$, in (b)), so that most qubits have relaxed at least once during the DD sequence. Despite arising from entirely different experimental platforms, the same strong relaxation-induced detuning dependence is observed. Red points are experimental data; solid black lines are analytic predictions (Eq.~\eqref{eq:all0_v_and_phase_first_order_in_delta}), where $\tau_{\pi} = T = 112~\text{ns}$ in (a) and $\tau_{\pi} = T = 300~\text{ns}$ in (b), with $\Omega = \pi/\tau_{\pi}$ in both cases. Each point is extracted from a sinusoidal fit to a fringe of $|1\rangle$ state population ($\langle p_1 \rangle$) as a function of the phase of the final state readout $\pi/2$ pulse $\phi_R$, with $\phi_R \in [0, 2\pi]$, using the fit function $\langle p_1 \rangle = \frac{1}{2}(1 + v\cos[\Delta\phi + \phi_R]) + b$ with fit parameters $v$, $\Delta\phi$, and $b$: an 11-point fringe in (a), and a 37-point fringe in (b) with $\langle p_1 \rangle$ measured over 2000 repetitions at each $\phi_R$ (see Sec.~\ref{sec:all0_theory_intro}). The zero detuning point on the horizontal axes is estimated based on where the visibility is minimized. Panels (c) and (d) show one such fringe explicitly, for the Sr interferometer at $\delta/\Omega\approx -2.8\%$ and the transmon qubit at $\delta/\Omega\approx-21.0\%$, respectively; each fringe fit yields a single data point in both the visibility and phase plots comprising (a), for the Sr data, or (b), for the transmon data. Red points are experimental data, and solid lines are the result of sinusoidal fitting.
    }
    \label{fig:all0_response_data}
\end{figure}

This result captures only the leading-order effect of relaxation-induced detuning dependence, and does not account for other sources of visibility loss present in the experimental data (e.g., a spread in detuning error within the ensemble average, inefficiency in the final state readout, errors in the Rabi frequency of the drive, etc.) which are expected to further reduce the measured visibility beyond what Eq. \eqref{eq:all0_v_and_phase_first_order_in_delta} predicts. This calculation is covered in greater detail in Appendix \ref{sec:gate-based_interferometer_response_analytics}.

The experimentally measured phase shifts and visibilities for a 504 $X$-gate sequence in a $^{88}$Sr atom interferometer and a 200 $X$-gate sequence in a flux-tunable transmon qubit are shown in Fig.~\ref{fig:all0_response_data}(a) and (b), respectively, as a
function of detuning.
Each point in these plots is extracted from a fit to an interference fringe of the $|1\rangle$-state population as a function of readout $\pi/2$ pulse phase $\phi_R$. Representative fringes for each platform are shown in Fig.~\ref{fig:all0_response_data}(c) and (d).
Both systems operate in a regime where the duration of the full DD-sequence is long compared to the lifetime of the $\ket{1}$ state so that, during the sequence, it is expected that most qubits have undergone a relaxation event at least once.
Equation~\eqref{eq:all0_v_and_phase_first_order_in_delta} plotted as solid black lines in Fig.~\ref{fig:all0_response_data} shows good qualitative agreement with both sets of data across the full detuning range probed.
This same qualitative dependence of $v$ and $\Delta\phi$ on detuning, despite the significant physical differences between the two platforms, suggests that the effect may broadly appear in two-level systems under relaxation.

The data in Fig.~\ref{fig:all0_response_data}(a) was collected in a Sr atom interferometer. A cloud of $^{88}$Sr atoms was first cooled to approximately 4~mK in a magneto-optical trap (MOT) operating on the $^1S_0 \leftrightarrow {}^1P_1$ transition. Prior to the CP sequence, the atoms were released from their trap and allowed to freely fall in vacuum. 
The CP sequence itself was applied using laser pulses resonant with the $^1S_0 \leftrightarrow {}^3P_1$ transition (with qubit transition frequency $\omega_q/2\pi\approx435$~THz, and energy relation time $T_1\approx21.3~\mu\text{s}$ \cite{ pucher_88sr_2025}), where each atom in the cloud was a qubit with $^1S_0$ as the
$\ket{0}$ state and $^3P_1$ as the $\ket{1}$ state.
We measured a Ramsey coherence time of $T_2^*\approx120~\text{ns}$, limited by the thermal spread of detunings across the atom cloud.
An M-Labs ARTIQ timing system generated the RF signals that drove an acousto-optic modulator which controlled the timing and phase of the laser pulses. Further details on the experimental apparatus are provided in \cite{wang2024robust}.

The data in Fig.~\ref{fig:all0_response_data}(b) was collected using a flux-tunable transmon qubit dispersively coupled to a readout resonator, measured in the hanger geometry and operated at 8~mK in a dilution refrigerator. The qubit was biased with zero current, where the transition frequency, measured in situ, was $\omega_q/2\pi = 4.898$~GHz, with an energy relaxation time $T_1 \approx 21~\mu\text{s}$ and a Ramsey coherence time $T_2^{*} \approx 33~\mu\text{s}$. Qubit drive and readout tones were delivered on separate cryogenic input lines, each attenuated and filtered at successive temperature stages. The qubit state was read out dispersively through its readout resonator at 7.321~GHz; the transmitted signal was amplified by a traveling-wave parametric amplifier (TWPA) at the mixing-chamber stage, followed by a 4--8~GHz high-electron-mobility-transistor (HEMT) amplifier at 4~K, with further amplification at room temperature. Control and readout pulses were synthesized directly at microwave frequencies by a Radio-Frequency System-on-Chip (RFSoC) ZCU216 field-programmable gate array (FPGA) control system running the QICK firmware~\cite{stefanazzi2022qick}. The system transmits and receives analog microwave signals but performs all waveform generation and readout demodulation digitally in real time.

\begin{figure}[hbt!]
    \centering
    \includegraphics[scale = 1]{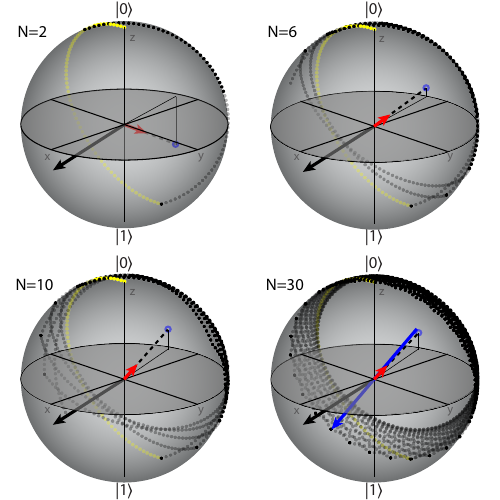}
    \caption{Bloch sphere OBE simulation (neglecting Rabi frequency and detuning inhomogeneity effects) showing the buildup of coherence from a sequence of $N$ repeating $X$-gate pulses in the driving field reference frame. The dots indicate the distribution taken by the decayed qubits. Yellow dots are specific to the arc produced by the most recent pulse (black rotation vector). The red vector is the resulting ``cumulative state," representing the density matrix of the set of qubits that have undergone decay. For visual clarity, this vector is extended by a dotted line to the blue circle which is, in this example, the intersection with the back side of the Bloch sphere in all images. In the $N=30$ Bloch sphere, the blue arrow represents the $\hat{\eta}$ axis which results from the combined pulse and dead time rotation vectors from Eq.~(\ref{combined_vector}). Anomalous black dots can be seen at the vertices between two arcs. These are many overlapping samples created during the dead times between pulses.}
    \label{fig:CP_sim_buildup}
\end{figure}\textbf{}

Both platforms implemented the same CP sequence structure shown in Fig. \ref{fig:all0_circuit_diagram}: the Sr interferometer operated with $T = \tau_\pi = 112$~ns ($\Omega/2\pi \approx 4.46$~MHz) and $T' = T/4$, while the transmon qubit operated with $T = \tau_\pi = 300$~ns ($\Omega/2\pi \approx 1.7$~MHz) and $T' = T/2$. To first order in $\delta$, the qubit response is insensitive to the particular choice of $T'$ (see Appendix \ref{sec:gate-based_interferometer_response_analytics}). In each case, a common detuning $\delta$ was introduced by offsetting the frequency of the drive field from the qubit resonance. Each resulting fringe was fit to the functional form given in the caption of Fig.~\ref{fig:all0_response_data}, with baseline $b$, visibility $v$, and phase shift $\Delta\phi$ as free parameters; quoted uncertainties on $v$ and $\Delta\phi$ are the $1\sigma$ uncertainties obtained from the covariance matrix of each fit.

The data plotted in Fig.~\ref{fig:all0_response_data} for both the Sr and transmon platforms is an average over many qubits. In the Sr interferometer, this averaging occurred naturally from the readout mechanism: each shot of the experiment interrogated all atoms in the cloud simultaneously, where $\langle p_1 \rangle$ was inferred from fluorescence photon counts averaged across all atoms in the cloud.
In the transmon qubit system, only a single qubit was addressed per shot of the experiment,
but the same averaging occurred naturally from the repeated shots performed on that qubit,
with $\langle p_1 \rangle$ inferred from the shot-averaged demodulated readout signal, projected onto a fixed quadrature and rescaled using reference levels for the ground and excited states.
The Sr cloud typically consisted of millions of atoms per shot, with multiple fluorescence photons collected per atom, while the transmon qubit data was collected using 2000
repetitions of the $\langle p_1 \rangle$ measurement for each value of detuning $\delta$ and $\phi_R$. Despite these differences in how $\langle p_1 \rangle$ was measured between these two systems, both platforms exhibited the same relaxation-induced dependence of $v$ and $\Delta\phi$ on detuning, as shown in Fig.~\ref{fig:all0_response_data}.

The buildup of the detuning-dependent coherence can be visualized on the Bloch sphere. Figure~\ref{fig:CP_sim_buildup} shows the distribution of qubit states produced after different numbers of $\pi$ pulses, for a sequence in which the dead time between consecutive $\pi$ pulses is equal to the $\pi$ pulse duration. Different points on the Bloch sphere correspond to qubits that decayed at different times during the sequence and were subsequently evolved through the remaining pulses and free-evolution intervals. The points were generated using an approximate semi-numerical model of dynamics under the  optical Bloch equations (OBE), which describe driven two-level dynamics in the presence of excited-state decay \cite{scully1997quantum,metcalf1999laser}. In this model, decay events are treated as instantaneous resets to the ground state that interrupt otherwise coherent evolution. Namely, for each decay time the qubit is initialized in $\ket{0}$, then propagated coherently from that point to the end of the sequence. The pulse sequence is divided into 65 possible decay times for each $\pi$ pulse and for each free evolution interval. The coherent propagation is computed by setting the decay rate to zero after the reset. This procedure approximates the distribution of final states by treating each sampled decay time as the qubit’s most recent reset before the end of the sequence; for visualization, the decay times are sampled uniformly and the undecayed population is omitted. Although the plotted distributions assume a single decay event per qubit, additional decay events would reset the qubit back to $\ket{0}$, erasing information about its previous state and causing it to re-emerge on a later arc. Thus, for the purpose of describing the final state, it is sufficient to track the qubit only after its last decay. By pulse 30, a clear structure has formed in the distribution.

\begin{figure}[hbt!]
    \centering
    \includegraphics[scale = 1]{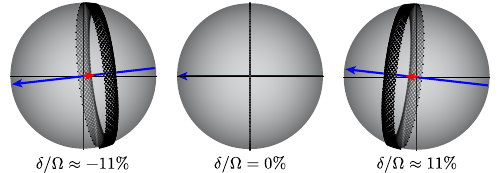}
    \caption{OBE simulated Bloch spheres (neglecting Rabi frequency and detuning inhomogeneity effects) showing the final distribution of qubit states from previously decayed qubits after 50 $X$-gate pulses, prior to the final dead time, $T'$, and readout pulse. The Bloch spheres shown are for a small negative, zero, and positive detuning. This figure highlights how the direction of the cumulative state vector (red arrow) is highly sensitive to the driving field detuning. Note the flip in direction of the cumulative state vector, which is associated with the flip of the measured phase shift by $\pi$ in Fig. \ref{fig:all0_response_data}, as the detuning changes sign. As Fig. \ref{fig:CP_sim_buildup}, the blue arrow represents the $\hat{\eta}$ axis.}
    \label{fig:Bloch_sphere_detuning}
\end{figure}

This structure can be understood geometrically as a band on the surface of the Bloch sphere with approximate rotational symmetry around a particular axis $\hat{\eta}$. This axis is determined by the repeated rotations applied during the sequence.  From Eq.~(\ref{eqn:sequenceGates}), we see that the qubits undergo repeated, alternating periods of free evolution (of duration $T$) and $X$ gates.  We denote the Bloch sphere rotation undergone during each period of free evolution as $\hat{R}_{\hat{\delta}}[\left|\delta \right| T]$, where $\hat{\delta}  = \text{sign}(\delta) \mathbf{z}$ is the rotation axis set by the sign of the detuning and $\left|\delta \right| T$ is the rotation angle.  We denote the Bloch sphere rotation undergone during each $X$ gate as $\hat{R}_{\hat{\Omega}}[\sqrt{\Omega^2 + \delta^2} \tau_\pi]$, where the rotation axis $\hat{\Omega}$ accounts for modifications from nonzero detunings (see Fig. \ref{fig:CP_arc}(b)) and the rotation angle is $\sqrt{\Omega^2 + \delta^2} \tau_\pi$.  The direction of $\hat{\Omega}$ is defined by the vector sum of $\left|\Omega \right|\mathbf{x}$ and $\left|\delta \right| \hat{\delta}$. The combined action of these two rotations can be expressed as a single rotation by an angle $\theta$ about an axis $\hat{\eta}$, with $\theta$ and $\hat{\eta}$ determined by $\Omega$, $\delta$, $T$, and $\tau_\pi$:

\begin{equation}
    \hat{R}_{\hat{\eta}}[\theta] = \hat{R}_{\hat{\Omega}} \left[\sqrt{\Omega^2 + \delta^2} \tau_\pi\right] \hat{R}_{\hat{\delta}}\left[\left|\delta \right| T\right].
\label{combined_vector}
\end{equation}

\noindent Points on the Bloch sphere generated by decay during a given pulse or free evolution interval are subsequently acted upon by repeated applications of this effective rotation.  As the number of these rotations increases, these points are carried around the axis $\hat{\eta}$, producing the aforementioned band-like structure. Figure~\ref{fig:Bloch_sphere_detuning} displays the structure for three representative detuning values (before the readout $\pi/2$ pulse rotation), showing the detuning-dependent coherence that leads to the interference indicated in Fig.~\ref{fig:all0_response_data}.  Further discussion of this geometric picture can be found in Appendix \ref{Appendix:Geometric_Picture}.

\subsection{Comparing Relaxation-Induced Detuning Sensitivity Across Other DD Sequences}

In Sec.~\ref{sec:all0_theory_intro}, we considered the CP sequence (a train of $\pi$ pulses with fixed phase, which we denote $X, X, X, ...$)
in the limit that the sequence duration is long relative to $T_1$,
where relaxation causes the interferometer visibility to be small near $\delta=0$ and to grow as $|\delta|$ increases. This behavior, however, is not generic to all DD sequences, but instead depends sensitively on the choice of pulse phases. As an example, in a previous work we observed markedly different behavior in a related sequence with pulse phases alternating between $\pi/2$ and $-\pi/2$, which, following the $X$-gate notation introduced above, we denote $-Y,Y,-Y,Y, ...$ . This alternating-phase sequence, which has also been used to suppress the effect of pulse-area errors \cite{bishof2013optical}, instead yields a relaxation-induced interferometer visibility at $\delta=0$ that is large 
\cite{wang2024robust}.
In this section, we build on this observation in two ways. First, rather than restricting the analysis to the
regime of Sec.~\ref{sec:all0_theory_intro}, where the duration of the DD sequence is long compared to $T_1$,
we track the emergence of this relaxation-induced coherence directly, by measuring the visibility of the $-Y,Y,-Y,Y, ...$ sequence as a function of pulse number.  Moreover, we provide an intuitive explanation for the contrasting behaviors between the $-Y,Y,-Y,Y, ...$ and CP sequences. Second, we use numerical simulations to examine several other common DD sequences (XY-4, XY-8, KDD, UR-8 \cite{souza_robust_2011, ryan_robust_2010, genov_arbitrarily_2017}), and compute how strongly the phase shift between the $\ket{0}$ and $\ket{1}$ states  depend on small detunings across a range of relaxation rates $\gamma = 1/T_1$.

\begin{figure}[hbt!]
    \centering
    \includegraphics[scale = 1]{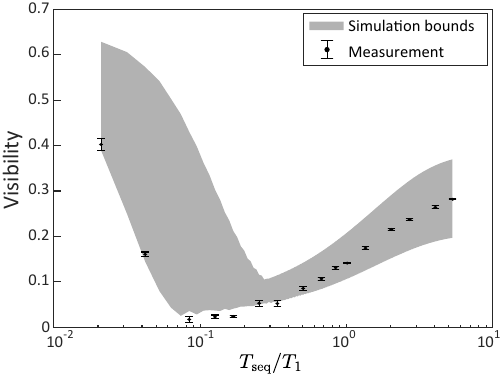}
    \caption{Dissipation-enabled emergence of coherence in a pulse sequence of alternating $-Y$ and $Y$-gate operations for the Sr atom interferometer. Experimental data shows a rapid decline in visibility when increasing the sequence duration, $T_{\text{seq}}$, by adding more pulses, due to pulse inefficiencies from detuning inhomogeneities and Rabi frequency inhomogeneities. Beyond $T_{\text{seq}}/T_1 \approx 10^{-1}$, visibility begins to re-emerge due to population that has undergone decay, with visibility continuing to rise beyond the $T_1$ decay time of $21.3~\mu s$. The error bars for the experimental data represent the standard error in the fitted visibility. Following the approach from \cite{wang2024robust}, simulation bounds are determined from the uncertainty bounds of each experimental parameter. The experimental parameters used in the simulation are the peak Rabi frequency at center of the Gaussian laser beam ($\Omega_{\text{peak}} = 5.07 \pm 0.18$ MHz), the ratio of the Gaussian beam waist size to r.m.s. atom cloud radius ($w_0/\sigma = 3.06 \pm 0.36$), detuning ($\delta = 0 \pm 0.10$ MHz), and detuning inhomogeneity Gaussian standard deviation ($\Delta \nu = 0.88 \pm 0.05$ MHz) \cite{wang2024robust}. Note that due to the driving field spatial inhomogeneity and detuning inhomogeneity, the optimal peak Rabi frequency $\Omega_{\text{peak}}$ was slightly larger than the  Rabi frequency corresponding to $\pi/\tau_{\pi}$, $\Omega = 4.46$ MHz. The plotted simulation bounds account for imperfections in the detection process, as described in Appendix \ref{Appendix:Simulation correction}.
    \label{fig:CP_emergence}
    }
\end{figure}

Figure~\ref{fig:CP_emergence} plots the visibility versus $T_{\text{seq}}/T_1$ for the $-Y,Y,-Y,Y,... $ sequence with $\delta=0$, measured using the Sr atom interferometry platform described in Sec.~\ref{sec:all0_theory_intro}, where the DD sequence duration $T_{\text{seq}}$ is varied by adding additional pulses to the sequence while holding $\tau_\pi=T=112~\text{ns}$ fixed.
The sequence begins with a zero-phase $\pi/2$ pulse, followed by the alternating $-Y,Y$ $\pi$ pulses, and concludes with a readout $\pi/2$ pulse. For small $\pi$ pulse numbers, the visibility initially decayed due to pulse infidelities \cite{wang2024robust} arising from the ensemble distribution of Doppler-shift-induced detunings due to the finite temperature, as well as from Rabi frequency inhomogeneities caused by the finite size of the atom cloud relative to the driving laser beam. As the $\pi$ pulse number increased, however, the visibility progressively recovered as the total sequence duration approached and exceeded $T_1$, in qualitative agreement with simulations based on the OBE. The simulations follow the same approach as those in \cite{wang2024robust}.

\begin{figure*}[t]
    \centering
    \includegraphics[width=\textwidth]{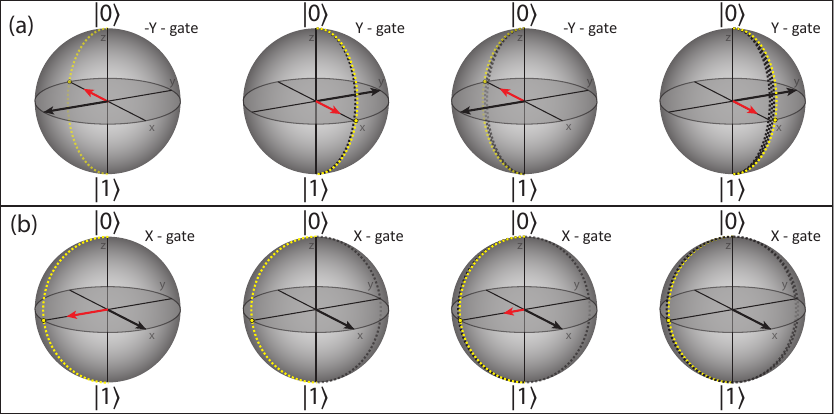}
    \caption{Bloch sphere visual aid depicting the distribution of decayed qubit states during: (a) a sequence of repeating $-Y$, $Y$-gate operations, and (b) a sequence of repeating $X$-gate operations. The yellow distribution arc is the distribution created by the most recently applied pulse (black vector). The cumulative state (red vector) persists in (a) due to the previous distribution arcs overlapping with newly created ones and is suppressed in (b) due to equally opposing distribution arcs.}
    \label{fig:Multipanel_CP_arc}
\end{figure*}

Figure \ref{fig:Multipanel_CP_arc}(a) provides a Bloch sphere depiction of the dissipation-enabled coherence buildup for this sequence.  During the first $\pi$ pulse, qubits that decay generate an arc at azimuthal angle $\pi$ on the Bloch sphere. The subsequent pulse rotates this arc to azimuthal angle $0$ while simultaneously generating a new arc of freshly decayed qubits at the same azimuthal angle. The two arcs are therefore overlapped, and their contributions add. Repeating this pattern throughout the sequence causes decay events from successive pulses to accumulate onto a common arc, reinforcing the coherence with each pulse. In the limit where the total sequence duration greatly exceeds the excited-state lifetime, the qubits that have undergone decay dominate the final readout signal.

By contrast, for the CP sequence (see Fig.~\ref{fig:Multipanel_CP_arc}(b)), the arc produced by each pulse lies at azimuthal angle $-\pi/2$ on the Bloch sphere. However, successive $\pi$ pulses invert the orientation of the existing arc from the previous pulse, causing the contributions from neighboring pulses to be anti-aligned and cancel. In the absence of detuning, this cancellation is complete, rendering the ensemble visibility zero. Introducing a finite detuning $\delta$ breaks this cancellation.  Namely, as discussed above, the detuning imparts an additional phase rotation between pulses and tilts the rotation axis during each pulse, producing an average state whose magnitude grows proportionally to $\left|\delta \right|$. 

The emergence of a large coherence at $\delta=0$ for the alternating $-Y,Y$ sequence can alternatively be understood by analogy with a CP sequence with a finite effective detuning.  A central role of detuning in the CP sequence is to change the relative phase between the drive field and the qubit coherence by $\delta T$ during the dead time between consecutive $\pi$ pulses.  In the alternating $-Y,Y$ sequence, the drive field phase itself changes by $\pi$ between neighboring pulses, producing an analogous relative phase advance.  This corresponds to an effective detuning $\delta_{\text{eff}} = \pi/T$.  Though the effective detuning does not have an identical quantitative effect as an actual detuning, which also modifies the rotation axis during pulses, the analogy is useful because the alternating drive phase plays a qualitatively similar role to the detuning-induced phase advance between pulses.

\begin{figure}
\includegraphics[width=3.40in]{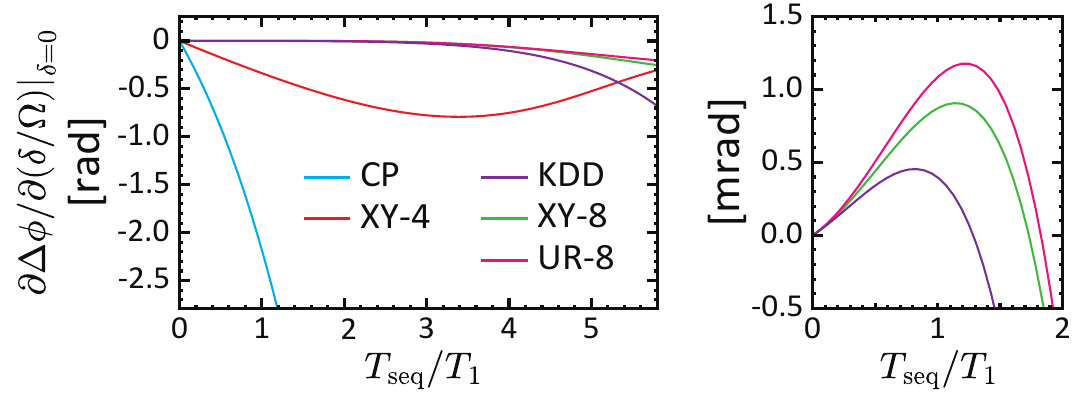}
\caption{
Simulated sensitivity of five common DD sequences  (CP, XY-4, XY-8, KDD, and UR-8) to relaxation-induced detuning errors. For each sequence (40 $\pi$ pulses, total duration $T_{\text{seq}}\approx6.4~\mu\text{s}$, $\Omega = 2\pi\times6.25 ~ \text{MHz}$), we plot the slope of the interferometer phase shift with respect to detuning, $\partial\Delta\phi/\partial(\delta/\Omega)$, as a function of the $|1\rangle$-state relaxation rate $\gamma = 1/T_1$. All sequences are chosen to perfectly cancel detuning-dependent dephasing in the absence of relaxation ($\partial\Delta\phi/\partial\delta = 0$ at  $\gamma=0$), so any nonzero slope at finite $\gamma$ reflects a relaxation-induced breakdown of dynamical decoupling.
Both panels show the same simulated data, plotted as a function of $T_{\text{seq}}/T_1$. The right panel zooms in on the three best-performing sequences, KDD, XY-8, and UR-8, to show the extent to which their phase response diverges from zero.
The CP sequence is the most sensitive to relaxation, while XY-8 and UR-8 remain comparatively robust across the full range of $\gamma$ probed.
}
\label{fig:DD_sequence_comparison}
\end{figure}

We now turn to numerical simulations to explore the impact of relaxation across a broader set of DD sequences.
We restrict our attention to DD sequences with a fixed pulse timing, in which only the phase of each $\pi$ pulse is varied. This excludes sequences such as UDD \cite{uhrig_keeping_2007} and WDD \cite{hayes_reducing_2011}, for which the deadtime between pulses is modulated, as well as continuous dynamical decoupling sequences \cite{Cai2012Continuous, Timoney2011Continuous}, for which there is no deadtime between pulses at all.
In Fig. \ref{fig:DD_sequence_comparison}, we compare five such sequences: the CP sequence \cite{carr_effects_1954}, the XY-4 and XY-8 sequences \cite{gullion_new_1990}, the KDD sequence \cite{ryan_robust_2010, souza_robust_2011}, and the UR-8 sequence \cite{genov_arbitrarily_2017}.
Each simulated sequence consists of 40 $\pi$ pulses, with the phase pattern repeated to fill the 40 total pulse phases (e.g., the 5-pulse KNILL composite pulse is repeated 8 times, for 40 total $\pi$ pulses).
We simulate with a Rabi frequency of $\Omega = 6.25~\text{MHz}$ (giving a $\pi$ pulse duration $\tau_\pi = \pi/\Omega = 80~\text{ns}$), the deadtime between mirror $\pi$ pulses of $T = \tau_\pi = 80~\text{ns}$, and a total sequence duration of $T_{\text{seq}} \approx 6.4~\mu\text{s}$, all held fixed across the five DD sequences. 
We scan the relaxation rate $\gamma = 1/T_1$ for each of these sequences to map out a continuum of relaxation regimes, from $1/T_1 = 0$, where no qubit decay occurs, up through large $1/T_1$, where effectively all qubits decay during the interferometer sequence, which is the same limit considered in the analysis of Sec.~\ref{sec:all0_theory_intro}.

Because no DD sequence can decouple a qubit from $T_1$ losses, relaxation reduces the visibility $v$ even for sequences with no detuning-induced dephasing at all; a perfect DD sequence in the absence of relaxation would give $v=1$ and $\Delta\phi=0$. To isolate the detuning-dependent effect specifically, we instead benchmark each sequence by the slope of the phase shift $\partial\Delta\phi/\partial(\delta/\Omega)$ in the small-$\delta$ limit, evaluated for different values of $\gamma$ and a constant value of $\Omega$. For each value of $\gamma$, we evaluate $v$ and $\Delta\phi$ over 11 detuning values between $-10~\text{kHz}$ and $+10~\text{kHz}$, using the same fitting routine described in the caption of Fig.~\ref{fig:all0_response_data} (but applied to numerically generated, rather than experimentally measured, excited-state populations), and fit the resulting $\Delta\phi[\delta]$ values to a line. The extracted slope is plotted against $\gamma$ in Fig.~\ref{fig:DD_sequence_comparison}.
We denote the deadtime between the first $\pi/2$ pulse and the first $\pi$ pulse as $T'$. For instantaneous pulses, the optimal choice would simply be $T' = T/2$; the finite duration of real pulses \cite{lang2019nonvanishing}, however, introduces a small residual detuning dependence in $\Delta\phi$ at $\gamma=0$ unless $T'$ is adjusted slightly away from $T/2$. In the experimental data of Fig.~\ref{fig:all0_response_data}, the atom interferometer used $T'=T/4$ and the qubit used $T'=T/2$ -- both close to, though not exactly, the value that fully cancels this residual dependence. We expect the resulting small deviation from perfect cancellation to be negligible compared to the relaxation-induced effects that are the focus of this section. Here, by contrast, we choose $T' \approx 14.54~\text{ns}$ so that every simulated sequence perfectly cancels the detuning-dependent phase shift when $\gamma=0$, isolating the effect of relaxation alone on each sequence's detuning sensitivity. We further assume no ensemble spread in detuning errors contributing to a single measurement of $\Delta\phi$, unlike an atom cloud, where each shot averages over the thermal spread of detunings across the atoms, or a qubit system, where one might expect averaging over a slightly different qubit detuning on each shot of the experiment. We also assume that the pulse amplitudes are chosen to exactly give the desired pulse area ($\pi$ for $\pi$ pulses, and $\pi/2$ for $\pi/2$ pulses).

The five sequences respond very differently to relaxation. The CP sequence is the most sensitive: $\partial\Delta\phi/\partial(\delta/\Omega)$ grows quickly with $\gamma$.
This divergence reflects the approximate analytic expression for the phase shift $\Delta\phi$ (Eq. \eqref{eq:all0_v_and_phase_first_order_in_delta}), which is evaluated in the $T_{\text{seq}}/T_1\rightarrow\infty$ limit and is a piecewise constant function with infinite slope at $\delta=0$.
The XY-8 and UR-8 sequences, by contrast, perform substantially better, with $\partial\Delta\phi/\partial\delta$ remaining small even as $\gamma$ increases.
Fig.~\ref{fig:DD_sequence_comparison} shows the same phase shift sensitivity curves, plotted as a function of $T_{\text{seq}}/T_1$. The right panel zooms in on the three best-performing sequences, KDD, XY-8, and UR-8, to show the extent to which their phase response diverges from zero for smaller values of $\gamma$.
The CP sequence, taken to the many-pulse limit, is the sequence underlying both the 504-pulse atom interferometer data and the 200-pulse flux-tunable qubit data shown in Fig.~\ref{fig:all0_response_data}, and gives rise to the strong relaxation-induced phase sensitivity observed there.  We note that all of the pulse sequences studied retain some detuning dependence for nonzero $\gamma$.  This dependence may be important to consider in quantum information and metrology applications requiring precise phase control even for sequences in which the effect is more suppressed, and even when the total sequence duration is significantly smaller than $T_1$.  Furthermore, our analysis indicates that relaxation-induced phase sensitivity to detuning may be an important factor in deciding which sequence is optimal for a given application.

\begin{figure}[hbt!]
    \centering
    \includegraphics[scale = 1]{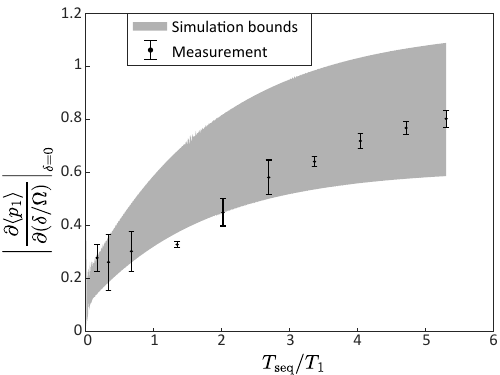}
    \caption{Experimental measurements, using the Sr atom interferometer, and OBE simulation showing the emergence of a detuning sensitive coherence for a CP sequence that strengthens beyond the $T_1$ relaxation time of the atoms. In both experiment and simulation, the atom cloud excited state population  $\langle p_1 \rangle$ is sampled five times over a detuning range of $\pm 200$ kHz.  The best fit slope $\left | \partial \langle p_1 \rangle /\partial (\delta / \Omega) \right |_{\delta=0}$ is plotted versus sequence duration. The error bars for the experimental data represent the standard error in the fitted slope. The simulation bounds are determined as in Fig. \ref{fig:CP_emergence}.}
\label{fig:CP_sense_buildup}
\end{figure}

We now present experimental data from the Sr atom interferometer showing how the the detuning senstivity of the readout grows as the DD sequence duration is increased relative to the $T_1$ time, for the case of a CP sequence. This measurement complements the simulated data in Fig.~\ref{fig:DD_sequence_comparison}, where the pulse sequence was held fixed and $T_1$ was varied. In the experiment, $T_1$ is fixed by the lifetime of the
$^3P_1$ state of Sr, so we instead varied $T_{\mathrm{seq}}$ with $T_1$ fixed. The pulse duration and interpulse spacing were held fixed at
$\tau_{\pi}=T=112~\mathrm{ns}$, while the number of $\pi$ pulses was varied to obtain different values of $T_{\mathrm{seq}}$.
For each sequence length, the phase of the final readout $\pi/2$ pulse was fixed at $\phi_R=\pi/2$, and the drive detuning was varied. The excited-state population $\langle p_1 \rangle$ was then fit to a line as a function of the normalized detuning $\delta/\Omega$. Figure~\ref{fig:CP_sense_buildup} plots the slope of this fit,
$\partial \langle p_1 \rangle /\partial(\delta/\Omega)$, at fixed $\phi_R=\pi/2$, as a function of the DD sequence duration $T_{\mathrm{seq}}$. Notably, this detuning dependence markedly increases as $T_{\mathrm{seq}}$ increases past $T_1$. The data show qualitative agreement
with OBE simulations that account for inhomogeneities in detuning and Rabi frequency across the ensemble.
The Sr interferometer data in Fig.~\ref{fig:all0_response_data}, which show the detuning-dependent phase shift and visibility, were collected with 504 $\pi$ pulses, corresponding to the rightmost data point in Fig.~\ref{fig:CP_sense_buildup}. Using Eqs.~\eqref{vis_def} and
\eqref{eq:all0_v_and_phase_first_order_in_delta} with $\phi_R=\pi/2$, the
expected asymptotic slope is $\frac{\partial \langle p_1 \rangle}{\partial(\delta/\Omega)} = \frac{3}{4}$,which is approximately the value approached by the experimental data in Fig.~\ref{fig:CP_sense_buildup} at long sequence duration.

\section{Dissipative Carr-Purcell Spectroscopy (DCPS)}\label{sec:DCPS}

The dissipation-induced sensitivity to detunings exhibited by the CP sequence (Fig.~\ref{fig:all0_response_data}) is a systematic effect that undermines the typical goal of DD sequences to suppress detuning-induced dephasing. On the other hand, this sensitivity is strong enough that we can turn the systematic into a signal. In this section, we use the CP sequence's dissipation-induced detuning dependence as the basis for a new spectroscopic technique which we named DCPS. 
The central advantage of DCPS over the widely used technique of Ramsey spectroscopy is that, within a single shot, DCPS suppresses the influence of detuning fluctuations at frequencies between $1/T_1$ and $1/T_2^*$, while retaining sensitivity to a constant detuning offset.
Qualitatively, this robustness to high-frequency noise arises because relaxation leads to an averaging over the detuning fluctuations, damping the qubit's response to fluctuations faster than the relaxation rate $\gamma=1/T_1$.

We now make this high-frequency noise suppression more quantitative. In Appendix~\ref{Appendix:PhaseNoise}, we construct an approximate analytical model for the response of the DCPS sequence to time-dependent drive field frequency noise, $\dot\phi_N[t]$, where $t$ denotes the time at which the noise is evaluated. $\phi_N[t]$ is the noise-induced phase difference between the drive field and the qubit, whose time derivative is the instantaneous frequency noise. 
Let $t=0$ denote the end of the sequence, immediately following the final $\pi/2$ pulse, and let $p_1[t_0]$ denote the $|1\rangle$-state population at this time for a qubit whose last relaxation event occurred at time $t_0$. Taking the sequence to be much longer than $T_1$, so that every qubit has relaxed at least once, the ensemble-averaged value of $p_1$ is then given by a weighted average over $t_0$:

\begin{equation}
\label{Eq:PopulationAverage}
\langle p_1 \rangle =
\int_{-\infty}^{0}
\!\!
dt_0
~
\left(\frac{\gamma}{2}\right)
e^{-\frac{\gamma}{2}(-t_0)}
p_1[t_0],
\end{equation}

\noindent
where the exponential weight reflects the probability that a given qubit's most recent relaxation event occurred at $t_0$, with more distant decays exponentially less likely. The factor of $1/2$ multiplying $\gamma$ reflects the approximation that the $\ket{0}$- and $\ket{1}$-state populations are approximately equal in the limit of long sequence duration with respect to $T_1$. As derived in Appendix~\ref{Appendix:PhaseNoise}, we evaluate $p_1[t_0]$ in the limit that $\left(T+\tau_{\pi}\right)\ll 1/\gamma$, for which Eq. \eqref{Eq:PopulationAverage} can be written as

\begin{equation}
\langle p_1 \rangle =
\frac{1}{2}
-
\frac{\gamma}{4}
\int_{-\infty}^{0}
\!\!
dt_0
\int_{t_0}^{0}dt_1
~
e^{-\frac{\gamma}{2}(-t_0)}
\sin\left[
A[t_1, t_0]\right]
\dot{\phi}_N[t_1]
\label{eqn:population_avg_main_text}
\end{equation}

\noindent
where $A[t_f,t_i] = \int_{t_i}^{t_f}dt_1\,\Omega[t_1]$ is the time-integrated Rabi frequency between $t_i$ and $t_f$. The time dependence of the Rabi frequency $\Omega[t_1]$ reflects the fact that the drive field is pulsed on and off, so that for square pulses, $\Omega[t_1]$ alternates between a set value during the pulses and zero during free evolution.

We can compute a transfer function $H[\omega_{\dot\phi}]$ that relates the power spectral density of the drive field's phase noise to the resulting r.m.s. error in a detuning measured using the DCPS sequence. In Appendix~\ref{Appendix:PhaseNoise}, we derive the form of $H[\omega_{\dot\phi}]$ by considering a single noise tone, $\dot\phi_N[t_1] = A_{\dot\phi}\cos[\omega_{\dot\phi}t_1+\phi_{\dot\phi}]$, substituting it into Eq.~\eqref{eqn:population_avg_main_text}, averaging in quadrature over all possible phases $\phi_{\dot\phi}$, and solving in the limit that the noise varies slowly in a single pulse cycle, $\dot\phi_N[t] \ll 1/(T+\tau_\pi)\;\forall t$, which gives

\begin{equation}\label{eq:transfer_function_H}
H[\omega_{\dot{\phi}}] = 
\sqrt{\frac{\gamma^2}{2\left(\gamma^2+4(\omega_{\dot{\phi}})^2\right)}}
\end{equation}

\noindent
the square root of a Lorentzian, with a full width at half maximum (FWHM) linewidth of $\sqrt{3}\gamma$. This Lorentzian suppression for noise frequencies higher than the relaxation rate, $\omega_{\dot\phi} > \gamma$, is precisely the averaging over high-frequency drive field phase noise described qualitatively at the beginning of this section.

For a full noise spectrum, described by the power spectral density $S_{\dot\phi}[\omega_{\dot\phi}]$, the total r.m.s. error in the measured detuning is then given by the quadrature sum of this transfer function against the noise spectrum,

\begin{equation}\label{eq:delta_rma}
\delta_{\text{RMS}}
=
\sqrt{
\int_{0}^{\infty}
d \omega_{\dot{\phi}}
\,S_{\dot{\phi}}[\omega_{\dot{\phi}}]
\left(H[\omega_{\dot{\phi}}]\right)^2
}
\end{equation}
\noindent
where $\delta_{\text{RMS}}$ is the r.m.s. error in the measured detuning.

We validated the noise averaging with numerical simulations and experimental data, and we compared the noise response to that of Ramsey spectroscopy. 
For the Ramsey data, we set the deadtime between $\pi/2$ pulses to be $T_R = T_2^* = 120~\text{ns}$. This time was limited by dephasing from the Doppler detuning spread of the Sr atom cloud and was much shorter than the $T_1$ time of our system ($\approx 21.3 ~ \mu\text{s}$).  By contrast, the sequences we implemented for DCPS were much longer ($\approx 113 ~ \mu \text{s}$). Consistent with the theoretical treatment above, operating in this $T_1 \gg T_2^*$ regime enables improved averaging of time-dependent frequency noise that varies slowly in comparison to the Ramsey pulse spacing but rapidly in comparison to the $T_1$ time.

We experimentally measured the transfer function $H[\omega_{\dot\phi}]$ at a given angular frequency $\omega_{\dot{\phi}} = 2 \pi f_{\dot{\phi}}$ by controllably introducing a laser phase Fourier component of the form

\begin{equation}
    \phi_N[t] = \alpha \sin\left[2\pi f_{\dot{\phi}} t + \phi_{\dot{\phi}}\right]
\label{freq_noise}
\end{equation}

\noindent In our experiments, we set $\alpha =  0.1$. This approximated frequency noise of amplitude $A_{\dot{\phi}} = 2 \pi \alpha f_{\dot{\phi}}$.  For 8 different values of $\phi_{\dot{\phi}}$, we extracted a measured noise response by taking the difference between the measured detuning with ($\alpha =  0.1$) and without ($\alpha =  0.0$) the artificially introduced noise.  The r.m.s. noise response of the measured detuning was then estimated by summing these individual noise responses in quadrature.  Further details of the measurement and numerical simulation of the transfer function can be found in Appendix \ref{Appendix:Alternative Sim}.

Figure \ref{fig:Susceptibility} shows the substantially suppressed noise susceptibility for DCPS as compared to Ramsey spectroscopy. Numerical simulations of the susceptibility are also shown.  Simulation and measurement are in qualitative agreement for Ramsey spectroscopy and are consistent with the estimated experimental noise floor for DCPS.  The noise floor is estimated based on the standard deviation of repeated measurements of the detuning with $\alpha=0$.

\begin{figure}[hbt!]
    \centering
    \includegraphics[scale = 1]{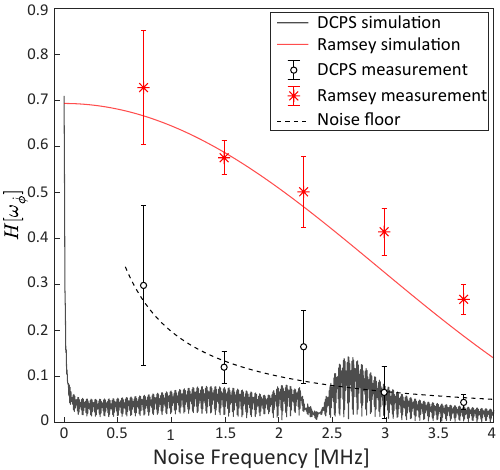}
    \caption{Simulated and experimentally measured frequency noise transfer function comparing DCPS to Ramsey spectroscopy. Error bars indicate the standard deviation from five independent measurements.  Measurements indicate the DCPS method is significantly less susceptible to MHz-scale noise, with transfer function measurements close to the noise floor of the experiment.}
    \label{fig:Susceptibility}
\end{figure}

\begin{figure}[hbt!]
    \centering
    \includegraphics[scale = 1]{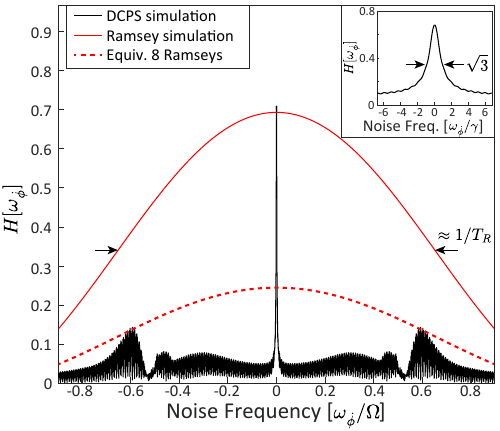}
    \caption{Double-sided simulated frequency noise transfer function highlighting the full frequency range of suppressed noise in DCPS over Ramsey spectroscopy. The inset is a magnified plot of the simulated DCPS noise transfer function which follows a square root Lorentzian curve with a full width at half maximum (FWHM) of $\sqrt{3} \gamma$. The values $\Omega = 2\pi \times 4.46$ MHz and $\gamma = 2\pi \times 7.47$ kHz were used to generate the shown curves. The Ramsey noise transfer function has a FWHM of $\sim 1/ T_R$, where $T_R$ is the dead time in the Ramsey sequence. An additional curve is plotted, which is the expected average of 8 independent Ramsey measurements. The number of Ramsey measurements in this curve was chosen to align with the highest DCPS transfer function value (above the decay frequency), and represents the fewest Ramsey measurements to reach an equivalent worst-case noise suppression from the DCPS method.}
    \label{fig:Susceptibility_sim_fullrange}
\end{figure}

To further highlight the noise suppression of DCPS, the double-sided noise susceptibility curve produced by simulation is provided in Fig. \ref{fig:Susceptibility_sim_fullrange}. The Ramsey simulation curve is approximately a sinc function with a FWHM of $\sim 1/T_R$. The simulations reproduce the $\sqrt{3}\gamma$ FWHM linewidth for DCPS of Eq. \eqref{eq:transfer_function_H}. An additional curve is included indicating the noise suppression equivalent of eight independent Ramsey measurements assuming a $1/\sqrt{M}$ averaging rate from $M$ measurements. This curve represents the minimum number of repeated Ramsey measurements needed to reach the worst case noise suppression from the DCPS method.

\begin{figure}[hbt!]
    \centering
    \includegraphics[scale = 1]{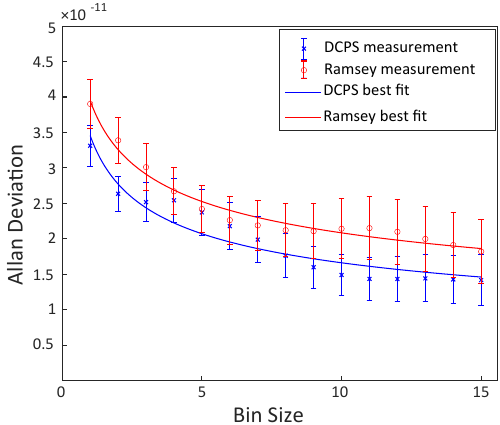}
    \caption{Comparison of Allan deviations for DCPS and Ramsey spectroscopy. For each spectroscopy method, 100 detuning measurements were collected. The Allan deviation is calculated using the overlapping bin method from \cite{riley2008handbook}. Error bars indicate the estimated standard deviation based on effective degrees of freedom under the approximation of white frequency noise. Each data set is fitted to the function of form $a x^{b}$, where x is the bin size and $a$ and $b$ are the fitting parameters. Ramsey is best fitted to $a=3.95 \times 10^{-11}$ and $b=-0.277$. DCPS is best fitted to $a=3.46 \times 10^{-11}$ and $b=-0.318$. Both the Ramsey and DCPS fitted $b$ parameters are consistent with a mix of both white and flicker frequency noise. ($b=-0.5$ corresponds to white frequency noise and $b = 0$ corresponds to flicker frequency noise.)}
    \label{fig:Allan Deviation}
\end{figure}

To further characterize the precision of DCPS, we compared its Allan deviation with that of Ramsey spectroscopy for our apparatus, with the results shown in Fig. \ref{fig:Allan Deviation}. We found that the Allan deviations for both methods are comparable. Although the DCPS method is slightly lower in deviation compared to Ramsey spectroscopy, this difference is not significant enough to be conclusive.

We conclude this section by considering the experimental circumstances under which DCPS might serve as a useful complement to Ramsey spectroscopy. First, the decay rate must be sufficiently large that decay occurs on experimentally relevant timescales. Second, DCPS is most likely to be useful when significant broadening limits the Ramsey coherence time, while technical noise, such as the frequency noise analyzed in this section, motivates averaging over a longer time than Ramsey spectroscopy can support under that broadening. This regime requires $T_1 \gg T_2^*$. The extended sequence duration needed for DCPS in this regime ($\gtrsim T_1$) is likely advantageous only when the pulse sequence itself does not limit the measurement repetition rate. This condition is satisfied in our Sr atom experiments, where the repetition rate is dominated by the multi-second timescale required to prepare the cold atoms, but it will often not be satisfied in other systems. Under this admittedly constrained combination of conditions, DCPS may be worth considering as a diagnostic tool. For instance, we have found it valuable as an efficient way to tune our atom-interferometer laser frequency to the center of the Doppler-shifted resonance of the atom cloud after release into free space. Ultimately, the utility of DCPS will need to be evaluated on a case-by-case basis across experimental platforms. Further studies of noise responses, systematic effects, and sources of drift remain topics for future work.

\section{Outlook}\label{sec:outlook}

In this work, we studied relaxation-induced detuning sensitivity in DD sequences with two different systems: a Sr atom interferometer and a superconducting transmon qubit. In both systems, the qubit was composed of an infinitely long-lived $\ket{0}$ state and a $\ket{1}$ state of finite lifetime $T_1$, and transitions between the two states were driven directly. We used this relaxation-induced detuning dependence to develop a new spectroscopic technique which we named DCPS. These results open several directions for future study.

Many of the possible extensions we discuss in the remainder of this section can be framed in terms of studying this effect in other qubit systems. Relevant candidates include trapped-ion qubits \cite{Haffner2008TrappedIons}, nitrogen-vacancy centers in diamond \cite{Doherty2013NitrogenVacancy}, Rydberg atoms \cite{Saffman2010RydbergAtoms}, molecular qubits \cite{Cornish2024MolecularSystem}, and freely falling atom systems using different internal states as qubit states than the ones studied here. These platforms can feature richer internal level structure than the effective two-level systems studied here, and in some of these systems the qubit resonance is driven indirectly via multi-photon transitions through one or more intermediate virtual states. Both of these features could have an interesting impact on the effect we have studied.

\textbf{Multi-photon gates: Relaxation only during the drive pulse.}
In some of the platforms described above, qubit transitions can also be driven via multi-photon operations, such as two-photon Raman transitions \cite{bergmann1998coherent, vitanov2017stimulated, kasevich1991atomic, leibfried2003quantum} or three-photon transitions \cite{panelli2025doppler, carman2025collinear, he2025coherent, ammenwerth2025realization, panelli2026microsecond}. In these transitions, the two qubit states are driven indirectly through two or more drive fields off-resonant with short-lived virtual states, rather than by a single field that directly couples the $\ket{0}$ and $\ket{1}$ states.
In these schemes, the two qubit states can be extremely long-lived (e.g. two hyperfine ground state sublevels driven by a two-photon Raman transition, or the $^1S_0$ ground state and $^3P_0$ clock state used in three-photon excitation schemes).
Relaxation instead arises because the drive briefly populates the short-lived virtual state during each pulse, from which the qubit can spontaneously decay directly. As a result, relaxation occurs mostly, or only, during the pulses themselves, rather than continuously throughout the sequence, as it does in the two systems studied in this paper. This feature could have an interesting impact on the effect studied in this paper.

\textbf{Effective $\ket{0}\rightarrow\ket{1}$ Relaxation}.
In this paper, we studied systems in which decay occurs only from the $|1\rangle$ state to the $|0\rangle$ state, but there are systems in which decay can also occur in the opposite direction. In the multi-photon schemes described above, the same virtual-state scattering mechanism which can lead to decay into the $\ket{0}$ state can also lead to decay into the $\ket{1}$ state, producing an effective $|0\rangle\rightarrow|1\rangle$ decay channel. This effective $|0\rangle\rightarrow|1\rangle$ relaxation can also arise directly from thermal excitation, where incoherent photons in the qubit's thermal environment drive residual population into the excited state \cite{Jin2015ThermalExcitationQubit}. The presence of this reversed relaxation channel could also meaningfully impact the effect described in this paper.

\textbf{Doppler shifts associated with relaxation recoil}. In solid-state qubit systems, a relaxation event does not shift the qubit's subsequent resonance frequency. In atomic or molecular qubits, however, a relaxation event involves the spontaneous emission of a photon in a random direction, which can impart a recoil kick to the atom or molecule and produce a Doppler shift, with the largest possible shift occurring when the emitted photon's wavevector is parallel or antiparallel to the wavevector of the light driving the transition. In tightly trapped atomic or molecular systems, in the limit of infinite trap stiffness, this recoil is absorbed by the trap, leaving the qubit's resonance frequency unshifted; in loosely trapped or free-fall systems, however, this recoil can shift the qubit's resonance frequency, which could have a meaningful impact on the dynamics studied here. Our free-fall Sr atom interferometer was, in principle, sensitive to this effect, but we operated with a Rabi frequency ($\Omega/2\pi \approx 4.46$~MHz) much larger than the maximum frequency shift associated with a single recoil event ($\sim10$~kHz). As a result, this effect was likely too small to be relevant in our measurements. The simulations and analysis described in Sec.~\ref{sec:mechanism_for_relaxation_induced_detuning_dependence_in_DD_sequences} do not account for this effect. In other free-fall or loosely trapped systems operated with smaller Rabi frequencies, however, this effect could become significant, and we leave a more detailed exploration of this effect for future study.

\textbf{Qubit-qubit interactions}. We neglected qubit-qubit interactions in the analysis presented in this paper. Even without any direct interaction between qubits, collective decay effects, such as superradiance, can produce effective qubit-qubit correlations \cite{Gross1982CollectiveDecay}, which could alter the assumptions underlying our analysis. Direct interparticle interactions offer another route to modify the dynamics \cite{Weiner1999Collisions, Reitz2022Dipole, fazio2025many}. Our atom cloud was likely too diffuse for either of these effects to have a significant impact on our measurements, but extending the current work to a regime where one or both of these effects become significant could have an impact on the relaxation-induced detuning effect studied here.

Each of these extensions (richer level structures, multi-photon drives with pulse-localized relaxation, $|0\rangle\rightarrow|1\rangle$ decay channels, recoil-induced Doppler shifts, and collective or interaction-driven decay) could modify the noise-suppression properties of DCPS. Evaluating the diagnostic utility of DCPS across each of these settings, and in systems with different levels of technical noise, could be a productive direction for future work.

\begin{acknowledgements}
\textit{Acknowledgments.}  T.R., H.J., and the Northwestern team acknowledge support by the U.S. Department of Energy, Office of Science, National Quantum Information Science Research Centers, Superconducting Quantum Materials and Systems Center (SQMS), under Contract No. 89243024CSC000002. Fermilab is operated by Fermi Forward Discovery Group, LLC under Contract No. 89243024CSC000002 with the U.S. Department of Energy, Office of Science, Office of High Energy Physics.  A portion of the work contributed by K.D., J.G., and S.S. was also supported by FermiForward Discovery Group, LLC under Contract No. 89243024CSC000002.  The Northwestern team was also supported by the National Science Foundation (Award Number 2409710), the Office of Naval Resarch (Contract Number N000142612236), the Gordon and Betty Moore Foundation (Grant ID GBMF7945.01), and the David and Lucile Packard Foundation (Fellowship for Science and Engineering).

Generative AI tools (Claude Opus 4.8 and Claude Sonnet 4.6 and 5.0, Anthropic; ChatGPT GPT-5.5, OpenAI) were used to assist with drafting and editing text, to suggest potentially relevant background literature, to suggest modifications to existing simulation code for improved speed, and to suggest a simplified analytical form for an integral. The tools were directed by the authors using author-supplied draft text, outlines, previous simulation code, and specific prompts. All AI-generated text and suggested revisions were reviewed and verified by the authors before inclusion, and every reference suggested by these tools was independently located and verified before citation. All revised code was carefully inspected and verified by comparison of the results against the original code.  The analytical form of the integral was verified independently. The authors take full responsibility for the content of the manuscript.
\end{acknowledgements}

\textit{Author Contributions.} The author list is arranged into groups of authors who contributed equally to the work, with names within each group arranged alphabetically, including the first two authors whose names are marked by a *.  J.G., K.J., and S.S. first observed the relaxation-induced detuning dependence of the Carr-Purcell DD sequence.  K.D. (lead), J.G. (lead), A.A. (supporting), K.J. (supporting), and S.S. (supporting) carried out data collection and analysis for the strontium atom platform.  H.J. and T.R. led data collection and analysis for the superconducting qubit.  K.D. led data collection for comparisons of DCPS and Ramsey spectroscopy. J.G. developed the gate-based and transfer function analysis, and K.D. developed the geometric theoretical analysis.  All authors contributed to the interpretation and validation of the results.  T.K. supervised the work.  K.D. (lead), J.G. (lead), K.J. (supporting), H.J. (supporting), T.K. (supporting), T.R. (supporting), and S.S. (supporting) prepared the initial manuscript draft.  All authors contributed to reviewing and revising the manuscript.

\appendix

\begin{widetext}
\section{Gate-based Estimate of Interferometer Response to Small Detuning Errors $\delta$}\label{sec:gate-based_interferometer_response_analytics}

Consider a Hamiltonian of the following form,
\begin{equation}\label{eq:hamiltonian_rotating_frame}
H = \hbar \begin{pmatrix}
\frac{\delta}{2} & \frac{\Omega[t]}{2}e^{i \phi_D[t]}
\\
\frac{\Omega[t]}{2}e^{-i \phi_D[t]} & -\frac{\delta}{2}
\end{pmatrix}
\end{equation}
\noindent
where $\Omega[t]$ is the Rabi frequency with which population is driven between the $\ket{0}$ and $\ket{1}$ states, $\delta = \omega_D - \omega_q$ is the detuning error between the drive frequency $\omega_D$ and the qubit frequency $\omega_q$ (such that $\delta$ is positive if the drive frequency is above resonance), and $\phi_D[t]$ is the phase of the drive field that couples the two states. The drive field amplitude, associated with the Rabi frequency $\Omega[t]$, and the drive field phase $\phi_D[t]$, are the control parameters available to us. For typical DD sequences, $\Omega[t]$ consists of square pulses with a piecewise-constant drive field phase $\phi_D[t]$. The Hamiltonian of Eq.~\eqref{eq:hamiltonian_rotating_frame} is written in a rotating frame; after applying the rotating-wave approximation to discard fast counter-rotating terms, what remains is a residual rotation at the detuning rate $\delta = \omega_D - \omega_q$, which appears above as the diagonal $\pm\delta/2$ terms \cite{metcalf1999laser}.
The qubit states in this frame can be expressed in this matrix form as

\begin{equation}
\begin{split}
| 0 \rangle &= \begin{pmatrix}
1 \\ 0
\end{pmatrix}
\\
| 1 \rangle &= \begin{pmatrix}
0 \\ 1
\end{pmatrix}
\end{split}
\end{equation}

For a constant Rabi frequency $\Omega[t] = \Omega$ and drive field phase $\phi_D[t] =
\phi_D$, the full time evolution operator for the system can be expressed as

\begin{equation}
U[t, \Omega, \phi] = \begin{pmatrix}
U_{11} & U_{12}
\\
U_{21} & U_{22}
\end{pmatrix}
\end{equation}

\noindent
with

\begin{equation}
\begin{split}
U_{11} &= 
\cos\left[\frac{1}{2}\Omega t \sqrt{1 + \left(\delta / \Omega\right)^2}\right]
- i \frac{\delta/\Omega}{\sqrt{1 + \left(\delta / \Omega\right)^2}}
\sin\left[\frac{1}{2}\Omega t \sqrt{1 + \left(\delta / \Omega\right)^2}\right]
\\
U_{12} &= 
-i e^{i \phi_D}
\frac{1}{\sqrt{1 + \left(\delta / \Omega\right)^2}}
\sin\left[\frac{1}{2}\Omega t \sqrt{1 + \left(\delta / \Omega\right)^2}\right]
\\
U_{21} &= 
-i e^{-i \phi_D}
\frac{1}{\sqrt{1 + \left(\delta / \Omega\right)^2}}
\sin\left[\frac{1}{2}\Omega t \sqrt{1 + \left(\delta / \Omega\right)^2}\right]
\\
U_{22} &= 
\cos\left[\frac{1}{2}\Omega t \sqrt{1 + \left(\delta / \Omega\right)^2}\right]
+ i \frac{\delta/\Omega}{\sqrt{1 + \left(\delta / \Omega\right)^2}}
\sin\left[\frac{1}{2}\Omega t \sqrt{1 + \left(\delta / \Omega\right)^2}\right]
\end{split}
\end{equation}

\noindent
where $t$ is the duration of the time evolution. Let us make the assumption that the
detuning is small, $\delta/\Omega \ll 1$. In this limit, we have

\begin{equation}
U[t, \Omega, \phi_D] = \begin{pmatrix}
\cos\left[\frac{1}{2}\Omega t\right] - i\frac{\delta}{\Omega}\sin\left[\frac{1}{2}\Omega t\right]
&
-i e^{i\phi_D} \sin\left[\frac{1}{2}\Omega t\right]
\\
-i e^{-i\phi_D} \sin\left[\frac{1}{2}\Omega t\right]
&
\cos\left[\frac{1}{2}\Omega t\right] + i\frac{\delta}{\Omega}\sin\left[\frac{1}{2}\Omega t\right]
\end{pmatrix}
+ \mathcal{O}\left[\delta^2\right]
\end{equation}

The gate-based operations indicated in Fig. \ref{fig:all0_circuit_diagram} can be expressed in terms of U as

\begin{equation}\label{eq:gate-based_matrix_definitions}
\begin{split}
X &= U\left[\frac{\pi}{\Omega}, \Omega, 0\right]
\\
X_t^{P} &= U\left[t, \Omega, 0\right]
\\
U_t^{F} &= U\left[t, 0, 0\right]
\\
R_{\mathbf{n}} &= U\left[\frac{\pi}{2a\Omega}, a\Omega, \phi_R\right]
\end{split}
\end{equation}

\noindent
where $X$ denotes a full $\pi$ pulse of duration $\tau_\pi = \pi/\Omega$ at drive phase $\phi_D = 0$, corresponding to the ideal $X$-gate operation applied throughout the DD sequence. A $Y$ gate would correspond to $\phi_D = -\pi/2$. $X_t^{P}$ denotes a partial $X$-gate: a drive of the same phase and Rabi frequency as $X$, but of arbitrary duration $t \leq \tau_\pi$, which we use below to describe a qubit that relaxes partway through a pulse. $U_t^{F}$ denotes free evolution for a duration $t$, with no drive field applied ($\Omega = 0$), representing the deadtime between pulses in the DD sequence. Finally, $R_{\mathbf{n}}$ denotes the final $\pi/2$ rotation used to read out the qubit state, with drive phase $\phi_R$ setting the axis $\mathbf{n}$ of this rotation on the equatorial plane of the Bloch sphere.
The final $\pi/2$ pulse operation can be performed by halving the pulse duration and keeping the same drive amplitude as the $\pi$ pulses, or by keeping the duration the same as a $\pi$ pulse and instead halving the drive amplitude; both choices produce the same net $\pi/2$ rotation. The $^{88}$Sr atom interferometer data (Fig. \ref{fig:all0_response_data}(a)) was collected using the former convention, i.e., $a = 1$ in the definition of $R_{\mathbf{n}}$ in Eq. \eqref{eq:gate-based_matrix_definitions}, while the superconducting transmon qubit data (Fig. \ref{fig:all0_response_data}(b)) was collected using the latter convention, i.e. $a = 1/2$. To capture both cases in the analysis that follows, we calculate for general $a$.

The state $|\psi\rangle$ of a qubit that has undergone a relaxation event at some point during the DD sequence, evaluated just before measurement (see Fig.~\ref{fig:all0_circuit_diagram}(a)), can be written as
\begin{equation}
|\psi\rangle = 
\begin{cases}
R_{\mathbf{n}}U_{T'}^{F}
\overbrace{X U_{T}^{F} X \cdots U_{T}^{F} X}^{\text{N X-gates}} U_{T}^{F}X^{P}_{\tau}|0\rangle,
&
\text{intra-pulse decay}
\\
R_{\mathbf{n}}U_{T'}^{F}
\underbrace{X U_{T}^{F} X \cdots U_{T}^{F} X}_{\text{N X-gates}} U_{\tau}^{F}|0\rangle,
&
\text{inter-pulse decay}
\end{cases}
\end{equation}

\noindent
where, following the relaxation event, the qubit is subject to $N$ subsequent $X$-gate operations before the final readout rotation $R_{\mathbf{n}}$. The relaxation event resets the qubit to $|0\rangle$ and the operator acting immediately after $|0\rangle$ ($X_\tau^P$ for intra-pulse decay, or $U_\tau^F$ for inter-pulse decay) captures only the remaining duration $\tau$ of whichever pulse or deadtime interval was underway at the moment of decay. All subsequent operators in the sequence then proceed with their full, unmodified durations.

Consider a measurement of the $|1\rangle$-state population of $|\psi\rangle$. We define the projection operator onto this state as

\begin{equation}
P_1 = | 1 \rangle \langle 1 |
\end{equation}

\noindent
so that the total population in the $|1\rangle$ state, $p_1$, is given by

\begin{equation}\label{eq:p1_in_terms_of_projection_P1}
p_1 = \langle \psi | P_1 | \psi \rangle
\end{equation}

Using the matrices of Eq.~\eqref{eq:gate-based_matrix_definitions}, $p_1$ can be written as a piecewise function depending on whether the qubit decayed during a pulse or during free evolution, and on the parity of the number of subsequent $X$-gate operations to which it is subject before measurement,

\begin{equation}\label{eq:p1_piecewise}
p_1 = \begin{cases}
p_1^{\text{driven, even}}, & \text{intra-pulse decay, } N \text{ even}
\\
p_1^{\text{free, even}}, & \text{inter-pulse decay, } N \text{ even}
\\
p_1^{\text{driven, odd}}, & \text{intra-pulse decay, } N \text{ odd}
\\
p_1^{\text{free, odd}}, & \text{inter-pulse decay, } N \text{ odd}
\end{cases}
\end{equation}

\noindent
with

\begin{equation}
\begin{split}
p_1^{\text{driven, even}} &= \frac{1}{2}\left(1+\sin\left[\Omega \tau\right]\cos\left[\phi_R\right]\right)
+\frac{\delta}{2\Omega}\left(-1 + \cos\left[\Omega \tau\right] - \left(\frac{1}{a}+\Omega T'\right)\sin\left[\Omega \tau\right]\right)\sin\left[\phi_R\right] + \mathcal{O}\left[\delta^2\right]
\\
p_1^{\text{free, even}} &= \frac{1}{2} + \mathcal{O}\left[\delta^2\right]
\\
p_1^{\text{driven, odd}} &= \frac{1}{2}\left(1-\sin\left[\Omega \tau\right]\cos\left[\phi_R\right]\right)
+\frac{\delta}{2\Omega}\left(-1 - \cos\left[\Omega \tau\right] + \left(\frac{1}{a} - \Omega \left(T-T'\right)\right)\sin\left[\Omega \tau\right]\right)\sin\left[\phi_R\right] + \mathcal{O}\left[\delta^2\right]
\\
p_1^{\text{free, odd}} &= \frac{1}{2} - \frac{\delta}{\Omega}\sin\left[\phi_R \right] + \mathcal{O}\left[\delta^2\right]
\end{split}
\end{equation}

\noindent
where the superscript `driven' or `free' denotes whether the qubit was in the middle of a pulse or in the middle of a free-evolution interval at the moment of relaxation, and the superscript `even' or `odd' denotes the parity of $N$, the number of subsequent $X$-gate operations applied after the relaxation event and before the final readout rotation $R_{\mathbf{n}}$. 
The reason $p_1$ depends only on these four conditions (rather than, for example, the absolute value of $N$) has to do with the structure of a single two-pulse cycle, consisting of a deadtime, a pulse, another deadtime, and a second pulse. To first order in $\delta$, the time evolution associated with one full two-pulse cycle is proportional to the identity, up to an overall phase:

\begin{equation}
X U_{T}^{F} X U_{T}^{F} = \begin{pmatrix}
-1 & 0 
\\
0 & -1
\end{pmatrix}
+\mathcal{O}[\delta^2]
\end{equation}

\noindent
Since this overall factor of $-1$ is a global phase, it has no effect on the measured population $p_1 = \langle\psi|P_1|\psi\rangle$. As a result, applying any additional \emph{complete} two-pulse cycles after a relaxation event leaves $p_1$ unchanged to first order in $\delta$. All that determines $p_1$, then, is where within a single two-pulse cycle the relaxation event itself occurs, indexed by two conditions: the parity of $N$, which identifies whether the decay occurs in the first or second half of a two pulse cycle, and whether the decay occurs during free propagation or during a pulse, together giving the four cases of Eq. \eqref{eq:p1_piecewise}.

Let us now consider averaging over all possible times a qubit could decay during the DD sequence, weighted by the distribution of expected decay times. We assume the DD sequence is long enough compared to the qubit relaxation time that every qubit in the system decays at least once during the sequence, and we further assume the characteristic relaxation time is long compared to the duration of two $X$-gate operations, so that a relaxed qubit is equally likely to be subject to an odd or an even number of subsequent $X$-gate operations following its decay. This is the approximation $2(T+\tau_\pi) \ll
1/\gamma$.

Because $p_1$ depends on the decay time only through which of the four cases of
Eq. \eqref{eq:p1_piecewise} applies, and each of these cases recurs identically every two-pulse cycle, it suffices to average over decay times within a single representative two-pulse cycle rather than over the full $N$-pulse sequence. We perform this average in two steps: first, we average over the possible decay times within each of the four piecewise cases of Eq. \eqref{eq:p1_piecewise} individually, evaluating each as an integral over $\tau$; second, we combine these four averages, weighting each by the fraction of the two-pulse
cycle corresponding to that case.

\begin{equation}\label{eq:p_1_values_after_averaging_over_tau}
\begin{split}
\left\langle p_1^{\text{driven, even}}\right\rangle_{\tau} = \left(\frac{\pi}{\Omega}\right)^{-1}\int_{0}^{\pi/\Omega} d\tau
\;
p_1^{\text{driven, even}}
&=
\left(\frac{1}{2} + \frac{1}{\pi}\cos\left[\phi_R\right]\right)
-
\left(\frac{T' \delta}{\pi} + \left(\frac{1}{2} + \frac{1}{a \pi}\right)\frac{\delta}{\Omega}\right)\sin\left[\phi_R\right] + \mathcal{O}\left[\delta^2\right]
\\
\left\langle p_1^{\text{free, even}} \right\rangle_{\tau} = 
\left(T\right)^{-1}\int_{0}^{T}d\tau
\;
p_1^{\text{free, even}}
&= \frac{1}{2} + \mathcal{O}\left[\delta^2\right]
\\
\left\langle p_1^{\text{driven, odd}}\right\rangle_{\tau} = \left(\frac{\pi}{\Omega}\right)^{-1}\int_{0}^{\pi/\Omega} d\tau
\;
p_1^{\text{driven, odd}}
&=
\left(\frac{1}{2} - \frac{1}{\pi}\cos\left[\phi_R\right]\right)
-
\left(\frac{\left(T-T'\right) \delta}{\pi} + \left(\frac{1}{2} - \frac{1}{a \pi}\right)\frac{\delta}{\Omega}\right)\sin\left[\phi_R\right] + \mathcal{O}\left[\delta^2\right]
\\
\left\langle p_1^{\text{free, odd}} \right\rangle_{\tau} = 
\left(T\right)^{-1}\int_{0}^{T}d\tau
\;
p_1^{\text{free, odd}}
&=
\frac{1}{2}
-\frac{\delta}{\Omega}\sin\left[\phi_R\right]
+\mathcal{O}\left[\delta^2\right]
\end{split}
\end{equation}

The value of $p_1$ averaged over all possible decay times is then a weighted sum of these four terms,

\begin{equation}\label{eq:time_averaged_p_1}
\begin{split}
\left\langle p_1\right\rangle =& 
\left(\frac{\tau_{\pi}}{2\left(T + \tau_{\pi}\right)}\right)
\left\langle p_1^{\text{driven, even}}\right\rangle_{\tau}
+
\left(\frac{T}{2\left(T + \tau_{\pi}\right)}\right)
\left\langle p_1^{\text{free, even}} \right\rangle_{\tau}
\\
&
+
\left(\frac{\tau_{\pi}}{2\left(T + \tau_{\pi}\right)}\right)
\left\langle p_1^{\text{driven, odd}}\right\rangle_{\tau}
+
\left(\frac{T}{2\left(T + \tau_{\pi}\right)}\right)
\left\langle p_1^{\text{free, odd}} \right\rangle_{\tau}
\\
=& \frac{1}{2} - \frac{2 T + \tau_{\pi}}{2\left(T+\tau_{\pi}\right)}\frac{\delta}{\Omega}\sin\left[\phi_R\right]
+\mathcal{O}\left[\delta^2\right]
\end{split}
\end{equation}

\noindent
where, for example, the weight $\tau_\pi/2(T+\tau_\pi)$ reflects the fraction of the two-pulse cycle during which the qubit is being driven, out of the four total intervals (a pulse, a deadtime, a pulse, and a deadtime) that make up the cycle. Writing Eq.~\eqref{eq:time_averaged_p_1} in the form of a typical interferometer fringe with phase shift $\Delta\phi$ and visibility $v$,

\begin{equation}
\langle p_1\rangle = \frac{1}{2}\left(1 + v \cos\left[\Delta \phi + \phi_R\right]\right)
+\mathcal{O}\left[\delta^2\right]
\end{equation}

\noindent
with $v$ defined to be positive, we can express the phase shift and visibility in terms of the detuning error as

\begin{equation}
\begin{split}
v & = \frac{2 T + \tau_{\pi}}{T+\tau_{\pi}}\left|\frac{\delta}{\Omega}\right|
\\
\Delta\phi & =
\pi + 
\begin{cases}
-\frac{\pi}{2}, & \delta > 0
\\
\frac{\pi}{2}, & \delta < 0
\end{cases}
\end{split}
\label{vis_equ}
\end{equation}

\noindent
which is Eq. \eqref{eq:all0_v_and_phase_first_order_in_delta} from the main text. Notably, the parameter $a$ (introduced earlier to distinguish the Sr interferometer's and transmon qubit's differing implementations of the final $\pi/2$ rotation) appears in the intermediate averaged terms (Eq. \eqref{eq:p_1_values_after_averaging_over_tau}) but cancels entirely once those terms are combined (Eq. \eqref{eq:time_averaged_p_1}), so that this final result for $v$ and $\Delta\phi$ is the same for both platforms. In Fig.~\ref{fig:all0_response_data}, both the $^{88}$Sr atom interferometer and the transmon qubit data were collected with $T=\tau_\pi$, for which this expression predicts a visibility that scales as $v \approx \frac{3}{2}\left|\delta/\Omega\right|$, shown as the solid black theory line in both panels.

\section{Response of Dissipative Carr-Purcell Spectroscopy to Detuning Noise}\label{Appendix:PhaseNoise}

Consider a Hamiltonian in a frame where the small time-dependent phase noise $\phi_N[t]$ is on the diagonal of $H$, but the piecewise-constant phase of the drive field $\phi_D[t]$ is on the off-diagonal,

\begin{equation}\label{eq:Hamiltonian_with_laser_frequency_noise}
H = \hbar \begin{pmatrix}
\epsilon \frac{1}{2}\dot{\phi}_N[t] & \frac{\Omega[t]}{2}e^{i \phi_D[t]}
\\
\frac{\Omega[t]}{2}e^{-i \phi_D[t]} & -\epsilon\frac{1}{2}\dot{\phi}_N[t]
\end{pmatrix}
\end{equation}

\noindent
which is essentially the Hamiltonian of Eq.~\eqref{eq:hamiltonian_rotating_frame}, but with a time-dependent detuning (setting $\dot{\phi}_N[t] = \delta$ recovers the Hamiltonian of Eq. \eqref{eq:hamiltonian_rotating_frame}). Here, $\epsilon$ is a bookkeeping parameter used to track orders of the expansion in the detuning noise. Expanding to first order in $\epsilon$ is equivalent to expanding to first order in the noise. Taking the drive field to have a constant phase, $\phi_D[t] = \phi_D$, for a given duration, the time evolution operator for the Hamiltonian of Eq. \eqref{eq:Hamiltonian_with_laser_frequency_noise} can be written, to first order in $\epsilon$, as

\begin{equation}\label{eq:appendix_B_U}
U\left[t_f, t_i, \phi_D\right] = \begin{pmatrix}
U_{11}
&
U_{12}
\\
U_{21}
&
U_{22}
\end{pmatrix}
\end{equation}

\noindent
where the matrix elements can be written as

\begin{equation}\label{eq:appendix_B_U_elements}
\begin{split}
U_{11} &= \cos\left[\frac{1}{2}A\left[t_f, t_i\right]\right]
-
\epsilon
\frac{i}{2} \int_{t_i}^{t_f}dt_1  Q_c\left[t_1, t_f, t_i\right]\dot{\phi}_N\left[t_1\right]
+\mathcal{O}\left[\epsilon^2\right]
\\
U_{12} &= e^{i \phi_D}\left(
-i \sin\left[\frac{1}{2}A\left[t_f, t_i\right]\right]
+ \epsilon\frac{1}{2}\int_{t_i}^{t_f}dt_1  Q_s\left[t_1, t_f, t_i\right]\dot{\phi}_N\left[t_1\right]
\right)
+\mathcal{O}\left[\epsilon^2\right]
\\
U_{21} &= e^{-i \phi_D}
\left(
-i \sin\left[\frac{1}{2}A\left[t_f, t_i\right]\right]
- \epsilon\frac{1}{2}\int_{t_i}^{t_f}dt_1  Q_s\left[t_1, t_f, t_i\right]\dot{\phi}_N\left[t_1\right]
\right)
+\mathcal{O}\left[\epsilon^2\right]
\\
U_{22} &= \cos\left[\frac{1}{2}A\left[t_f, t_i\right]\right]
+\epsilon
\frac{i}{2} \int_{t_i}^{t_f}dt_1  Q_c\left[t_1, t_f, t_i\right]\dot{\phi}_N\left[t_1\right]
+\mathcal{O}\left[\epsilon^2\right]\\
\end{split}
\end{equation}

\noindent
and where

\begin{equation}
A[t_f, t_i] = \int_{t_i}^{t_f}dt_1 \Omega[t_1]
\end{equation}

\noindent
is the pulse area accumulated between an initial time $t_i$ and a final time $t_f$, and the functions

\begin{equation}\label{eq:Q_functions}
\begin{split}
Q_c\left[t_1, t_f, t_i\right] &= \cos\left[
\frac{1}{2}A[t_f, t_1]
-
\frac{1}{2}A[t_1, t_i]
\right]
\\
Q_s\left[t_1, t_f, t_i\right] &= \sin\left[
\frac{1}{2}A[t_f, t_1]
-
\frac{1}{2}A[t_1, t_i]
\right]
\end{split}
\end{equation}

\noindent
are related to the frequency noise transfer functions.

Now consider an interferometer sequence consisting of evolution under an arbitrary Rabi frequency from time $t=t_0<0$ to time $t=0$, with a constant drive phase $\phi_D=0$, followed by an instantaneous shift of the drive phase to $\pi/2$, and a final $\pi/2$ pulse from $t=0$ to $t=\tau_\pi/2$ (e.g., a CP sequence with a readout phase of $\phi_R = \pi/2$). Let us apply this time evolution operator to a qubit that begins, at time $t_0$, in the $|0\rangle$ state. The state of the qubit after the final $\pi/2$ pulse can be written as

\begin{equation}\label{eq:appendix_B_psi}
|\psi\rangle = 
U\left[\frac{\tau_{\pi}}{2}, 0, \frac{\pi}{2} \right]
U\left[0, t_0, 0\right]
|0\rangle
\end{equation}

\noindent
where the $U[0, t_0, 0]$ operator is capturing the impact of the rest of the CP sequence (time evolution from time $t_0$ to $0$) after a qubit has relaxed to state $\ket{0}$, and the $U[\tau_{\pi}/2, 0, \pi.2]$ operator is capturing the impact of the final readout $\pi/2$ pulse.

The excited state population $p_1$ can be written to first order in $\epsilon$ using Eqs. \eqref{eq:p1_in_terms_of_projection_P1}, \eqref{eq:appendix_B_psi}, \eqref{eq:appendix_B_U}, and \eqref{eq:appendix_B_U_elements}, as

\begin{equation}\label{eq:p1_equation_1}
\begin{split}
p_1 = &
\frac{1}{2}\left(
1 - \cos\left[
A[0, t_0]
\right]
\cos\left[
A[\tau_{\pi}/2, 0]\right]
\right)
\\
&
+ \epsilon
\Bigg(
-
\sin\left[A[0, t_0]\right]
\int_{0}^{\tau_{\pi}/2}
\!\!\!
dt_1
~
\left(
\frac{1}{2}
Q_s\left[t_1, \tau_{\pi}/2, 0\right]
\cos\left[\frac{1}{2}A[\tau_{\pi}/2, 0]\right]
+
\frac{1}{2}
Q_c\left[t_1, \tau_{\pi}/2, 0\right]
\sin\left[\frac{1}{2}A[\tau_{\pi}/2, 0]\right]
\right)
\dot{\phi}_N[t_1]
\\
&
\;\;\;\;\;\;\;\;
+
\sin\left[A[\tau_{\pi}/2, 0]\right]
\int_{t_0}^{0}
\!\!\!
dt_1
~
\left(
\frac{1}{2}
Q_s\left[t_1, 0, t_0\right]
\cos\left[\frac{1}{2}A[0, t_0]\right]
-
\frac{1}{2}
Q_c\left[t_1, 0, t_0\right]
\sin\left[\frac{1}{2}A[0, t_0]\right]
\right)
\dot{\phi}_N[t_1]
\Bigg)
\\
&
+\mathcal{O}\left[\epsilon^2\right]
\end{split}
\end{equation}

Now using the identities

\begin{equation}
\begin{split}
A\left[\tau_{\pi}/2, 0\right] &= \frac{\pi}{2}
\\
A\left[t_3, t_2\right] + A\left[t_2, t_1\right] &= A\left[t_3, t_1\right]
\\
A[t_2, t_1] &= - A[t_1, t_2]
\end{split}
\end{equation}

\noindent and using Eq. \eqref{eq:Q_functions},  Eq. \eqref{eq:p1_equation_1} can be written in the form

\begin{equation}\label{eq:p1_equation_2}
p_1 = \frac{1}{2}
-
\epsilon
\left(
\underbrace{
\sin\left[A[0, t_0]\right]
\int_{0}^{\tau_{\pi}/2}
\!\!
dt_1
\frac{1}{2}
\sin\left[
A[\tau_{\pi}/2, t_1]\right]
\dot{\phi}_N[t_1]
}_{\text{Term 1}}
+
\underbrace{
\int_{t_0}^{0}dt_1
\frac{1}{2}
\sin\left[
A[t_1, t_0]\right]
\dot{\phi}_N[t_1]
}_{\text{Term 2}}
\right)
+
\mathcal{O}\left[\epsilon^2\right]
\end{equation}

In the limit that the DD sequence is long enough that every single qubit has decayed at least once during the sequence, the averaged value of $p_1$ (averaged over all decay times $t_0$) can be written as

\begin{equation}
\langle p_1 \rangle =
\int_{-\infty}^{0}
\!\!
dt_0
~
\left(\frac{\gamma}{2}\right)
e^{-\frac{\gamma}{2}(-t_0)}
p_1
\label{eqn:excited_state_population_decay_weighting}
\end{equation}

\noindent
where the factor of $1/2$ in the exponential probability distribution arises because, 
for times $t$ in the DD sequence much longer than $T_1$, only half of the qubit population is in the excited state to first order in $\delta$. Consider a function $\Omega[t]$ which consists of a set of evenly spaced $\pi$ pulses, of duration $\tau_\pi$,
separated by a deadtime $T$. In the limit that the pulse spacing is small compared to
$1/\gamma$, we have

\begin{equation}
\int_{-\infty}^{0}
\!\!
dt_0
~
\left(\frac{\gamma}{2}\right)
e^{-\frac{\gamma}{2}(-t_0)}
\sin\left[A[0, t_0]\right]
\rightarrow 0,
\;\;\;\;\;\;
(T+\tau_{\pi}) \ll 1/\gamma
\end{equation}

\noindent
since, over a two-pulse cycle, $\sin\left[A[0,t_0]\right]$ sweeps symmetrically from $0$ to $+1$ and back during one pulse and from $0$ to $-1$ and back during the next.
This oscillation averages to zero in the approximation that the two-pulse cycle time is short compared to the relaxation timescale
$1/\gamma$.
Therefore `Term 1' in Eq.~\eqref{eq:p1_equation_2} averages to zero, leaving only `Term 2' such that the $\ket{1}$ state population averaged over all decay times can be expressed as

\begin{equation}\label{eq:population_average_integral_expression}
\langle p_1 \rangle =
\frac{1}{2}
-
\epsilon
\frac{\gamma}{4}
\int_{-\infty}^{0}
\!\!
dt_0
\int_{t_0}^{0}dt_1
~
e^{-\frac{\gamma}{2}(-t_0)}
\sin\left[
A[t_1, t_0]\right]
\dot{\phi}_N[t_1]
\end{equation}

\noindent
which is Eq. \eqref{eqn:population_avg_main_text} of the main text.

We now look to approximate the integral of
Eq.~\eqref{eq:population_average_integral_expression} in the limit $2(T+\tau_\pi) \ll 1/\gamma$ and $2(T+\tau_\pi) \ll 1/\dot\phi_N[t]\;\;\forall t$ (i.e., $A[t_2,t_1]$ oscillates rapidly relative to both the relaxation time and to changes in the drive field's phase noise). We begin by assuming that $T'$, the deadtime between the last $\pi$ pulse and the final $\pi/2$ pulse, is small enough relative to $1/\gamma$ that a negligible fraction of qubits relax during this interval. In this limit, $\Omega[t]$ reprents a train of square pulses of duration $\tau_\pi$ separated by deadtimes of duration $T$, and can be written as

\begin{equation}
\Omega[t] = 
\Omega_0 \Theta\left[\tau_{\pi} - \text{Mod}\left[-t, T+\tau_{\pi}\right]\right]
\end{equation}

\noindent
where $\Omega_0$ is the Rabi frequency of a pulse, and $\Theta[x]$ is the Heaviside theta function. The integral over the pulse area from time $t_1$ to time $t_2$, $A[t_2,t_1]$,  can be written as

\begin{equation}\label{eq:exact_A}
A[t_2, t_1] = F[t_2] - F[t_1]
\end{equation}

\noindent
with

\begin{equation}
F[t] = \int dt ~ \Omega[t] =  -\Omega_0 \left(
\tau_{\pi}\left\lfloor \frac{-t}{T+\tau_{\pi}} \right\rfloor
+\text{Min}\left[\text{Mod}\left[-t, T+\tau_{\pi}\right],\tau_{\pi}\right]
\right)
\end{equation}

$A[t_2,t_1]$ has both a slowly varying component, set by the average pulse duty cycle, and
rapidly oscillating features associated with the individual pulses within each cycle. To facilitate the integration of Eq. \eqref{eq:population_average_integral_expression} for particular choices of $\dot{\phi}_N[t]$, we approximate $A[t_2,t_1]$ by its slowly varying component alone, effectively smoothing over
the fast, pulse-to-pulse oscillations (see Fig. \ref{fig:smooth_A_approximation}),

\begin{equation}\label{eq:temporally_averaged_A}
A[t_1, t_0] \approx \frac{\tau_{\pi}}{\tau_{\pi}+T}\Omega_0 \left(t_1-t_0\right)
\end{equation}

Because this temporally smoothed approximation of $A[t_1, t_0]$ does not automatically preserve the exact
value of the double time-integral over a single two-pulse cycle (of duration
$2(T+\tau_\pi)$, the same cycle considered in
Appendix~\ref{sec:gate-based_interferometer_response_analytics}), we introduce a
correction factor $B$ so that the integral evaluated using the temporally smoothed approximation to $A[t_1, t_0]$
agrees with the exact integral of Eq. \eqref{eq:population_average_integral_expression} over a single two-pulse cycle. Solving for B in the equation

\begin{equation}
\int_{-2\left(T+\tau_{\pi}\right)}^{0}dt_0
\int_{t_0}^{0}dt_1
~
\sin\left[
A[t_1, t_0]\right]
=
B
\int_{-2\left(T+\tau_{\pi}\right)}^{0}dt_0
\int_{t_0}^{0}dt_1
~
\sin\left[
\frac{\tau_{\pi}}{\tau_{\pi}+T}\Omega_0 \left(t_1-t_0\right)
\right]
\end{equation}

\noindent
gives us

\begin{equation}
B = \frac{\tau_{\pi}(2T+\tau_{\pi})}{(T+\tau_{\pi})^2}
\end{equation}

Therefore, in the limit that $\dot\phi_N[t]$ varies slowly over the duration of a single 2-pulse cycle, the integral expression

\begin{equation}\label{eq:population_average_integral_expression_simplified}
\langle p_1 \rangle \approx
\frac{1}{2}
-
\epsilon
\frac{\gamma}{4}
\frac{\tau_{\pi}(2T+\tau_{\pi})}{(T+\tau_{\pi})^2}
\int_{-\infty}^{0}
\!\!
dt_0
\int_{t_0}^{0}dt_1
~
e^{-\frac{\gamma}{2}(-t_0)}
\sin\left[
\frac{\tau_{\pi}}{\tau_{\pi}+T}\Omega_0 \left(t_1-t_0\right)
\right]
\dot{\phi}_N[t_1]
\end{equation}

\noindent
gives the same result as the integral expression of Eq. \eqref{eq:population_average_integral_expression}, but with an integrand that is much easier to integrate for arbitrary $\dot{\phi}_N[t]$.

\begin{figure}
    \centering
    \includegraphics[width=3.4in]{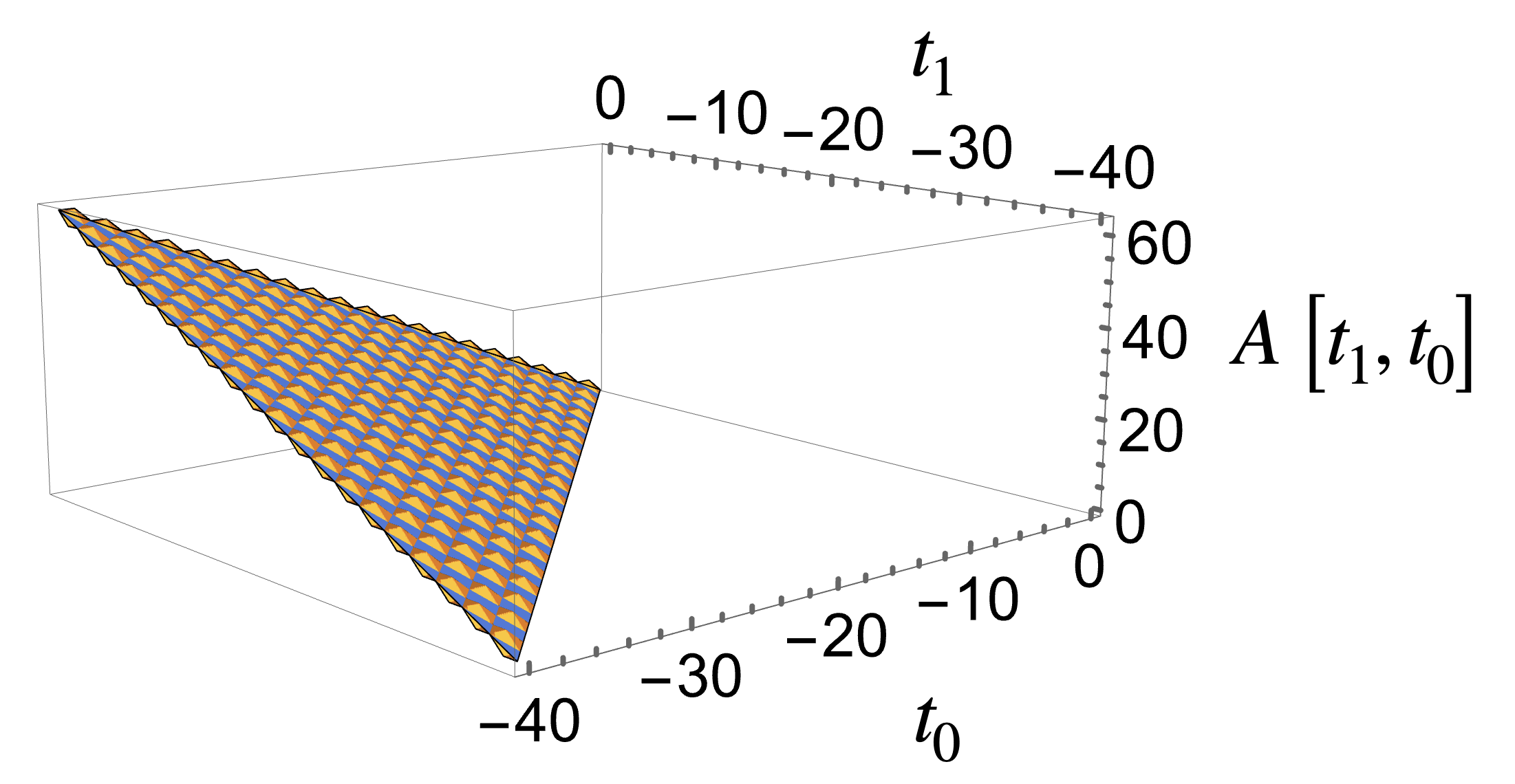}
    \caption{The temporally smoothed pulse area approximation used in Sec. \ref{Appendix:PhaseNoise}. $A[t_1, t_0]$ plotted using Eq. \eqref{eq:exact_A} (orange), and the approximate $A[t_1, t_0]$ of Eq. \eqref{eq:temporally_averaged_A} (blue) for $\Omega_0 = \pi$, $\tau_{\pi} = T = 1$}
    \label{fig:smooth_A_approximation}
\end{figure}

\subsection{Response to Constant Detuning}

To validate the result of Eq. \eqref{eq:population_average_integral_expression_simplified}, we consider the same scenario explored in Appendix \ref{sec:gate-based_interferometer_response_analytics}: a constant detuning between the
drive field and the qubit frequency.

\begin{equation}
\dot{\phi}_N[t] = \delta
\end{equation}

Solving the integrals of Eq. \eqref{eq:population_average_integral_expression_simplified} then gives

\begin{equation}
\langle p_1 \rangle \approx
\frac{1}{2}
-
\epsilon
\delta
\frac{2 \tau_{\pi}^2(2T+\tau_{\pi})\Omega_0}{\gamma^2(T+\tau_{\pi})^3 + 4 \tau_{\pi}^2(T+\tau_{\pi})\Omega_0^2}
\end{equation}

\noindent
which in the limit that $\gamma \ll 1/\left(T +\tau_{\pi} \right)$ becomes

\begin{equation}
\langle p_1 \rangle \approx
\frac{1}{2}
-
\epsilon
\frac{1}{2}
\frac{(2T+\tau_{\pi})}{T+\tau_{\pi}}
\frac{\delta}{\Omega_0}
\end{equation}

\noindent
in agreement with Eq.~\eqref{eq:time_averaged_p_1} for the case $\phi_R=\pi/2$. We can write this as

\begin{equation}
\langle p_1 \rangle \approx
\frac{1}{2}
+
\frac{\partial \langle p_1 \rangle}{\partial\delta}\big|_{\delta=0}
\delta
\end{equation}

\noindent
where

\begin{equation}\label{eq:p1_slope_from_delta}
\frac{\partial \langle p_1 \rangle}{\partial\delta}\big|_{\delta=0} = 
-
\epsilon
\frac{1}{2}
\frac{(2T+\tau_{\pi})}{(T+\tau_{\pi})\Omega_0}
\end{equation}

\noindent
is the sensitivity of the $p_1$ population to detuning. We can then invert this
relation to extract a measured detuning $\delta_{\text{measured}}$ between the drive field
and the qubit resonance from a measured value of $\langle p_1\rangle$,

\begin{equation}\label{eq:detuning_as_a_function_of_p1}
\delta_{\text{measured}} = \left(\frac{\partial \langle p_1 \rangle}{\partial\delta}\big|_{\delta=0}\right)^{-1}\left(\langle p_1 \rangle - \frac{1}{2}\right)
\end{equation}

\subsection{Response to an Oscillating Detuning}

Consider a driving field whose frequency is oscillating with amplitude $A_{\dot{\phi}}$ and frequency $\omega_{\dot{\phi}}$

\begin{equation}\label{eq:oscillating_frequency_noise}
\dot{\phi}_N[t_1] = A_{\dot{\phi}} \cos\left[\omega_{\dot{\phi}} t_1 + \phi_{\dot{\phi}}\right]
\end{equation}

Plugging Eq. \eqref{eq:oscillating_frequency_noise} in to Eq. \eqref{eq:population_average_integral_expression_simplified} gives us 

\begin{equation}
\langle p_1 \rangle \approx
\frac{1}{2}
-
A_{\dot{\phi}}\epsilon
\frac{2\gamma \tau_{\pi}^2 (2T+\tau_{\pi})\Omega_0
\left(
\gamma \cos[\phi_{\dot{\phi}}] + 2 \omega_{\dot{\phi}}\sin[\phi_{\dot{\phi}}]
\right)
}
{(T+\tau_{\pi})\left(\gamma^2(T+\tau_{\pi})^2 + 4 \tau_{\pi}^2\Omega_0^2\right)\left(\gamma^2 + 4 (\omega_{\dot{\phi}})^2\right)}
\end{equation}

\noindent
Then taking the limit $\gamma^2 \left(T+\tau_{\pi}\right)^2 \ll 1$, we have

\begin{equation}\label{eq:equation_for_p1_omega_phi}
\langle p_1 \rangle \approx
\frac{1}{2}
+
A_{\dot{\phi}}
\left(
\frac{\partial \langle p_1 \rangle}{\partial\delta}\big|_{\delta=0}\right)
\left(
\gamma \cos[\phi_{\dot{\phi}}] + 2 \omega_{\dot{\phi}}\sin[\phi_{\dot{\phi}}]
\right)
\frac{\gamma}{\gamma^2 + 4 (\omega_{\dot{\phi}})^2}
\end{equation}

\noindent
where the slope of the expected response to the $\ket{1}$ state population to constant drive field detunings $\delta$ is taken from Eq. \eqref{eq:p1_slope_from_delta}. 
For a single-frequency noise component, the r.m.s.\ shift in the measured $|1\rangle$-state
population away from $1/2$ can be written as an r.m.s.\ average over all possible phases
$\phi_{\dot\phi}$ of this noise component,

\begin{equation}
\left\langle
\langle p_1 \rangle - \frac{1}{2}
\right\rangle_{\phi_{\dot{\phi}}}
=
\sqrt{
\frac{1}{2\pi}
\int_0^{2\pi}d\phi_{\dot{\phi}}
\left(
\langle p_1 \rangle - \frac{1}{2}
\right)^2
}
\end{equation}

\noindent
where $\langle \cdot \rangle_{\phi_{\dot\phi}}$ denotes an r.m.s.\ average over possible noise phases. Using the value of $\langle p_1\rangle$ from Eq.~\eqref{eq:equation_for_p1_omega_phi}, this becomes

\begin{equation}\label{eq:rms_phase_averaged_population_response}
\left\langle
\langle p_1 \rangle - \frac{1}{2}
\right\rangle_{\phi_{\dot{\phi}}}
=
A_{\dot{\phi}}
\left(
\frac{\partial \langle p_1 \rangle}{\partial\delta}\big|_{\delta=0}
\right)
\sqrt{\frac{\gamma^2}{2\left(\gamma^2+4(\omega_{\dot{\phi}})^2\right)}}
\end{equation}

Because a fluctuation in the measured $|1\rangle$-state population due to time-varying drive field phase noise is indistinguishable, in a single measurement, from one caused by a constant detuning error, we can relate this r.m.s.\ population shift to an equivalent r.m.s.\ error in the measured detuning via Eq.~\eqref{eq:detuning_as_a_function_of_p1},

\begin{equation}\label{eq:single_tone_measured_delta_rms}
\langle \delta_{\text{measured}} \rangle_{\dot{\phi}}
=
\left(\frac{\partial \langle p_1 \rangle}{\partial\delta}\big|_{\delta=0}\right)^{-1}
\left\langle
\langle p_1 \rangle - \frac{1}{2}
\right\rangle_{\phi_{\dot{\phi}}}
= A_{\dot{\phi}} H[\omega_{\dot{\phi}}]
\end{equation}

\noindent

\noindent
where we have defined $H[\omega_{\dot\phi}]$ as a transfer function relating the drive field's frequency noise spectrum to an r.m.s.\ error in the measured detuning. For a spectrum of drive field noise amplitudes $A_{\dot\phi}[\omega_{\dot\phi}]$, the total r.m.s.\ error in the measured detuning is then given by a quadrature sum of this transfer function against the noise power spectrum,

\begin{equation}\label{eq:appendix_delta_rms}
\langle \delta_{\text{measured}} \rangle_{\dot{\phi}}
=
\sqrt{
\int_{0}^{\infty}
d \omega_{\dot{\phi}}
\left(A_{\dot{\phi}}[\omega_{\dot{\phi}}]\right)^2 
\left(H[\omega_{\dot{\phi}}]\right)^2
}
\end{equation}

\noindent
where $(A_{\dot\phi}[\omega_{\dot\phi}])^2$ is the power spectral density of the
drive field's frequency noise. Eq. \eqref{eq:appendix_delta_rms} is Eq. \eqref{eq:delta_rma} of the main text.

Using Eq. \eqref{eq:single_tone_measured_delta_rms}, together with Eq. \eqref{eq:rms_phase_averaged_population_response} gives the shape of the transfer function,

\begin{equation}
H[\omega_{\dot{\phi}}] = 
\sqrt{\frac{\gamma^2}{2\left(\gamma^2+4(\omega_{\dot{\phi}})^2\right)}}
\end{equation}

\noindent
the square root of a Lorentzian with FWHM linewidth $\sqrt{3}\gamma$, as shown in Fig.~\ref{fig:Susceptibility_sim_fullrange}. This is the key feature that makes the dissipative CP sequence useful as a spectroscopic tool: dissipation suppresses the response to drive field noise for frequencies above the relaxation rate $\gamma$, isolating its sensitivity to the constant detuning offset of the drive field frequency.

\section{Geometric Approach Derivation}

\label{Appendix:Geometric_Picture}

Here, we elaborate on the geometric picture of the relaxation-induced detuning dependence outlined in Sec. \ref{sec:mechanism_for_relaxation_induced_detuning_dependence_in_DD_sequences}. Let $|\psi(t)\rangle$ be a particular quantum state of a qubit after undergoing a decay event during a pulse to state $|0\rangle$ and being rotated by the remainder of the pulse for a time $t$. Assuming that decay events happen uniformly over time, we can define the density matrix corresponding to the ensemble of qubits that decayed during this pulse as

\begin{equation}
    \rho = \frac{1}{\tau_\pi} \int^{\tau_\pi}_0 |\psi(t) \rangle \langle \psi(t)| dt
\end{equation}

\noindent
We represent this density matrix as a Bloch vector, defined as

\begin{equation}
\vec{a}_0 = \begin{pmatrix}
\text{Re}[2\rho_{10}] \\ \text{Im}[2\rho_{10}] \\ 2\rho_{00} - 1
\end{pmatrix}
\end{equation}

\noindent
A visual depiction of this vector can be seen in Fig. \ref{fig:CP_arc_appendix}.

\begin{figure}[hbt!]
    \centering
    \includegraphics[scale = 1]{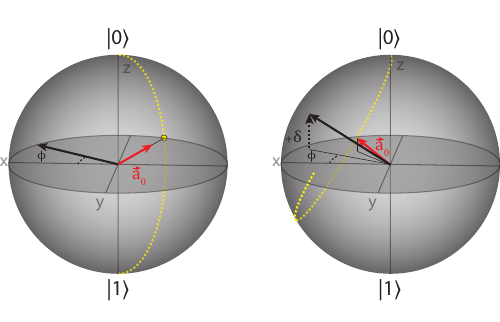}
    \caption{Two examples of the cumulative state from a single $\pi$ pulse in the Bloch sphere representation of a qubit in the driving field reference frame, with (right) and without (left) detuning. The solid black arrow at azimuthal angle $\phi$ represents the normalized rotation vector associated with the pulse.  The rotation angle associated with the pulse is $\sqrt{\Omega^2 + \delta^2} \tau_{\pi}$. The yellow dotted curve represents the distribution of qubit states after decay events to state $|0\rangle$ at different times during the pulse. Assuming that the decay events happen uniformly in time, the resulting cumulative state (representing the density matrix), $\vec{a}_{0}$, is given by the red arrow.}
    \label{fig:CP_arc_appendix}
\end{figure}

As noted in the main text, $\vec{a}_0$ will be subject to repeating $\hat{R}_{\hat{\delta}}\left[\left|\delta \right| T\right]$ and $\hat{R}_{\hat{\Omega}}\left[\sqrt{\Omega^2 + \delta^2} \tau_\pi\right]$ rotation operations from subsequent free evolution intervals and pulses, occurring about the axes $\hat{\delta}$ and $\hat{\Omega}$, respectively. These rotations can be treated as a single rotation of the form

\begin{equation}
    \hat{R}_{\hat{\eta}}[\theta]\vec{a}_0 = \hat{R}_{\hat{\Omega}}\left[\sqrt{\Omega^2 + \delta^2} \tau_\pi\right]\hat{R}_{\hat{\delta}}\left[|\delta| T\right]\vec{a}_0
\label{two_rotation}
\end{equation}

\noindent
After $n$ subsequent cycles of free evolution followed by a $\pi$ pulse, $\vec{a}_0$ is rotated to become

\begin{equation}
    \vec{a}_n = \left(\hat{R}_{\hat{\eta}}[\theta]\right)^n\vec{a}_0
\end{equation}

\noindent
Repeated rotations, $\left(\hat{R}_{\hat{\eta}}[\theta]\right)^k$ can be reduced to $\hat{R}_{\hat{\eta}}[k\theta]$ giving, 

\begin{equation}
    \vec{a}_n = \hat{R}_{\hat{\eta}}[n\theta]\vec{a}_0.
\label{nth rotation}
\end{equation}

\noindent
Thus, all $\vec{a}_n$ vectors for various $n$ lie at different multiples of $\theta$ about the unit vector $\hat{\eta}$. We argue below that the cumulative state vector that builds up after many pulses, which we approximate as the average of all $\vec{a}_n$, lies approximately parallel/anti-parallel to $\hat{\eta}$ for a sufficiently large number $N$ of pulses. To justify this, we can start by defining two orthogonal axes, $\mathbf{\hat{x}'}$ and $\mathbf{\hat{y}'}$, orthogonal to $\mathbf{\hat{\eta}}$, and write $\vec{a}_n$ in this new basis: $\vec{a}_n = a_{x',n}\mathbf{\hat{x}'} + a_{y',n}\mathbf{\hat{y}'} + a_{\eta,n}\mathbf{\hat{\eta}}$. Expressing $\vec{a}_n$ in the new basis allows us to reduce Eq. \ref{nth rotation} to

\begin{equation}
    \vec{a}_n = \left(a_{x',0} \cos[n\theta] - a_{y',0} \sin[n\theta]\right)\mathbf{\hat{x}'} + \left(a_{x',0} \sin[n\theta] + a_{y',0} \cos[n\theta]\right)\mathbf{\hat{y}'} + a_{\eta,0}\mathbf{\hat{\eta}}
\end{equation}

We now calculate an average $\langle \vec{a} \rangle_n$ over all arcs created during the sequence. For simplicity, we neglect any weighting by decay time of the type done in Eq. \eqref{eqn:excited_state_population_decay_weighting}. The components of $\langle \vec{a} \rangle_n$ are

\begin{equation}
\begin{split}
    \langle a_{x',n} \rangle = \frac{1}{N}\sum^{N-1}_{n = 0}{a_{x',0} \cos[n\theta] - a_{y',0} \sin[n\theta]}
\\
&
    = \frac{1}{N}\csc\left[\frac{\theta}{2}\right]\sin\left[\frac{N \theta}{2}\right]\left(a_{x',0}\cos\left[\frac{(N-1)\theta}{2}\right] + a_{y',0}\sin\left[\frac{(N-1)\theta}{2}\right]\right)
\\
    \langle a_{y',n} \rangle = \frac{1}{N}\sum^{N-1}_{n = 0}{a_{x',0} \sin[n\theta] + a_{y',0} \cos[n\theta]}
\\
&
    = \frac{1}{N}\csc{\left[\frac{\theta}{2}\right]}\sin\left[\frac{N \theta}{2}\right]\left(- a_{x',0}\sin\left[\frac{(N-1)\theta}{2}\right] + a_{y',0}\cos\left[\frac{(N-1)\theta}{2}\right]\right)
\\
    \langle a_{\eta,n} \rangle = \frac{1}{N}\sum^{N-1}_{n = 0}{a_{\eta,0}}
\\
&
    = a_{\eta,0}
\end{split}
\end{equation}

The $\langle a_{x',n} \rangle$ and $\langle a_{y',n} \rangle$ components converge to zero for $N \gg 1$ and $N \gg 1/|\theta|$, where $\theta$ is defined modulo $2 \pi$.  Once these conditions are met, the only nonzero component is $\langle a_{\eta,n} \rangle$, leaving $\langle \vec{a} \rangle_n = a_{\eta,0}\mathbf{\hat{\eta}}$. This shows that the cumulative state vector is a scaled $\mathbf{\hat{\eta}}$ vector related by the constant $a_{\eta,0}$.

\section{Further Details on Spectroscopy Measurements and Simulations}

\label{Appendix:Alternative Sim}

To experimentally measure the transfer function $H[\omega_{\dot{\phi}}]$ at a given angular frequency $\omega_{\dot{\phi}} = 2 \pi f_{\dot{\phi}}$, we approximated a noise Fourier component by modulating the phase of the drive laser such that a pulse beginning at time $t_1$ is set to phase of $\phi_N (t_1) = \alpha \sin[2\pi f_{\dot{\phi}} t_1 + \phi_{\dot{\phi}}]$, approximating frequency noise of amplitude $A_{\dot{\phi}} = 2 \pi \alpha f_{\dot{\phi}}$.  In the simulations shown in Figs. \ref{fig:Susceptibility} and \ref{fig:Susceptibility_sim_fullrange}, the driving phase is allowed to vary linearly across the driving pulse as a more accurate approximation of the frequency noise. Figure \ref{fig:Susceptibility_discrete_phase} shows alternative simulation curves that exactly models the experimental implementation.  The two simulation curves have the same qualitative behavior, reinforcing that the approximation of constant phase over the duration of each pulse captures the relevant qualitative features.

\begin{figure}[hbt!]
    \centering
    \includegraphics[scale = 1]{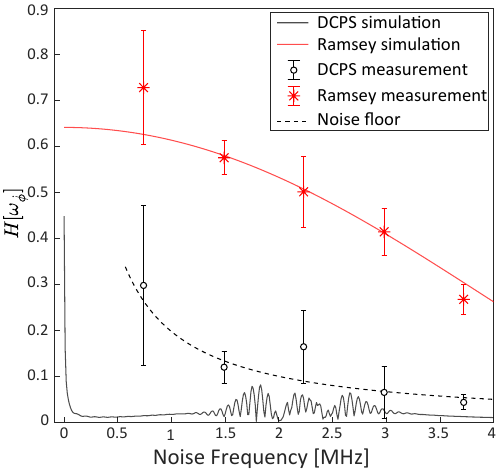}
    \caption{Alternative plot to Fig. \ref{fig:Susceptibility}. Here, the simulator treats the phase over each pulse as constant to better match the experimental implementation.}
    \label{fig:Susceptibility_discrete_phase}
\end{figure}

Both the Ramsey and DCPS spectroscopic measurements shared the same readout procedure once their respective sequences had concluded. A final (readout) $\pi/2$ pulse with $\pi/2$ phase leads to a linear relationship between laser frequency and excited state population ratio in the limit that the detuning is small. Under ideal conditions, the laser frequency corresponding to resonance can be extracted from the precise value of the laser frequency that gives a 50\% excited state population fraction. Our imaging process for extracting the excited state population fraction is known to underestimate the excited state population due to decay during the detection process. We circumvented this issue by repeating the sequences with a final $\pi/2$ pulse at $-\pi/2$ phase, giving the negative linear relationship between the laser frequency and the excited population fraction. The intersection of these two lines determines the resonance rather than the 50\% excited state population fraction. Each of our spectroscopic data points consisted of four independent runs of the sequence to create the two intersecting slopes: one pair of measurements with the laser frequency at $+/- 200$ kHz relative to the coarsely calibrated resonance with a final $\pi/2$ phase, and another pair at $+/- 200$ kHz relative to the  coarsely calibrated resonance with a final $-\pi/2$ phase.  These measurements then allowed us to determine the difference between the coarsely calibrated and actual resonance frequencies, which we refer to as the measured detuning in the main text.

For each chosen noise frequency, eight independent measurements of the detuning were made for different values of $\phi_{\dot{\phi}}$, equally spaced across $2\pi$, in order to approximate an rms average over the noise phase.  Specifically, the r.m.s. noise response is estimated as

\begin{equation}
    \Delta_{\text{RMS}}[f_{\dot{\phi}}] = \sqrt{\frac{1}{8}\sum_{i=0}^7{\left(\Delta \left[f_{\dot{\phi}}, \phi_{\dot{\phi}} = \frac{2 \pi i}{8} \right] \right)^2}}
\end{equation}

\noindent where $\Delta \left[f_{\dot{\phi}}, \phi_{\dot{\phi}} = \frac{2 \pi i}{8} \right]$ is the difference between the value of the detuning measured with introduced noise ($\alpha = 0.1$) and that measured with no introduced noise ($\alpha = 0$).  The measured value of the transfer function is then evaluated as

\begin{equation}
    H_{meas}[2 \pi f_{\dot{\phi}}] = \frac{\Delta_{\text{RMS}}}{2 \pi \alpha f_{\dot{\phi}}}
\end{equation}

\noindent with $\alpha=0.1$.

Since the DCPS sequence durations were much larger than $T_1$, the presence of the initial $\pi/2$ pulse did not significantly affect the readout signal, as decay made it unlikely for any memory of this pulse to persist to the end of the sequence.  For the DCPS measurements presented, this initial $\pi/2$ pulse was omitted. Pulse durations and spacings were the same as for the data described in Sec. \ref{sec:mechanism_for_relaxation_induced_detuning_dependence_in_DD_sequences}, with 504 $\pi$ pulses used.

\section{Further Details on Simulation Corrections}

\label{Appendix:Simulation correction}

In Figs. \ref{fig:CP_emergence} and \ref{fig:CP_sense_buildup}, the plotted simulation results for visibility versus number of pulses include a correction factor accounting for imperfections in the experimental detection process in order to more accurately compare with experimental measurements. Notably, in experiment, once the pulse sequence and readout pulse have concluded, we use a 461 nm laser pulse resonant with the $^1S_0 \leftrightarrow {}^1P_1$ transition to impart momentum onto the qubits in the $|0 \rangle$ state, leaving the qubits in the $|1 \rangle$ state undisturbed. After a subsequent drift time, this allows us to spatially distinguish qubits in state $|0 \rangle$ from those in state $|1 \rangle$ via fluorescence imaging \cite{wang2024robust}. For the data shown in these figures, the 461 nm pulse had a duration of 8 $\mu s$, after which time, only a fraction, $\beta =e^{-8/21.3}= 0.687$, of the excited state qubits will remain excited due to decay. We denote the experimentally measured excited state population as $p_1 = \beta p_1'$, where $p_1'$ is what the excited state population would be in the absence of decay during detection. The simulator does not include this decay effect, and it extracts a visibility value $v'$ by fitting $p_1'$ versus readout phase to the equation

\begin{equation}
    p_1' = \frac{1}{2} ( 1 + v' \cos[\Delta \phi + \phi_R]) + b'
    \label{equation:sim_fit}
\end{equation}

By contrast, we extract a visibility $v$ from our experiments using the experimental fit function stated in the Fig. \ref{fig:all0_response_data} caption.  By comparing this experimental fit function to \eqref{equation:sim_fit}, we can see that $v = \beta v'$.  Therefore, in the plots, we scale the simulated visibilies $v'$ by a factor of $\beta$.

\end{widetext}

\bibliography{DissipationDynamicalDecoupling.bib}

\end{document}